\documentclass[preprint,12p]{elsarticle}

\usepackage{amssymb, colortbl}

\journal{Astroparticle Physics}

\begin{document}
\begin{frontmatter}

\title{Dark Matter searches using gravitational wave
   bar detectors:  quark nuggets and newtorites }

\author[label2,label3]{M.~Bassan \corref{ronga}}
\author[label2,label6]{E.~Coccia}
\author[label3]{S.~D'Antonio}
\author[label2,label3]{V.~Fafone}
\author[label1]{G.~Giordano}
\author[label1]{A.~Marini}
\author[label3]{Y.~Minenkov}
\author[label2]{I.~Modena}
\author[label4]{G.V.~Pallottino}
\author[label1]{G.~Pizzella}
\author[label3]{A.~Rocchi}
\author[label1]{F.~Ronga \corref{ronga}}
\author[label5, label3]{M.~Visco}

\address[label1] {Istituto Nazionale di Fisica Nucleare - Laboratori Nazionali di Frascati, \\Via~
E.~Fermi,~40 - 00044 Frascati, Italy}
\address[label2] {Dipartimento di Fisica, Universit\`a di Tor Vergata, \\
Via~della~Ricerca~Scientifica,~ - 00133 Roma, Italy}
\address[label3] {Istituto Nazionale di Fisica Nucleare - Sezione Roma Tor Vergata, \\
Via~della~Ricerca~Scientifica,~ - 00133 Roma, Italy}
\address[label4] {Dipartimento di Fisica, Sapienza Universit\`a di Roma 
and Istituto Nazionale di Fisica Nucleare Ð Sezione Roma 1 --
 piazzale Aldo Moro 2,- 00185 Roma, Italy}
\address[label5] {Istituto di Astrofisica e Planetologia Spaziali - INAF \\
Via del Fosso del Cavaliere 100, 00133 Roma, Italy}
\address[label6] {Gran Sasso Science Institute, INFN, Viale F. Crispi 7 - 67100 L'Aquila, Italy}

\cortext[ronga]{Corresponding authors:  ronga@lnf.infn.it; ~ bassan@roma2.infn.it}
\date{today}

\begin{abstract}

Many experiments have searched  for supersymmetric  WIMP dark matter, with null results. This  may
suggest to look for more exotic  possibilities, for example compact ultra-dense quark nuggets,
widely discussed in literature with several different names.
Nuclearites are an example of candidate compact objects with atomic size cross section. After a
short discussion on  nuclearites, the result of  a nuclearite search with the gravitational wave bar
detectors Nautilus and Explorer is reported. The geometrical acceptance of the bar detectors is 19.5
$\rm m^2$ sr, that is smaller than that of other detectors used for similar searches. However, the
detection mechanism is  completely different and is more straightforward than in other detectors.
The experimental  limits we obtain are of  interest because, for nuclearites of mass less than
$10^{-5}$ g, we find a flux  smaller than that one predicted considering  nuclearites  as   dark
matter candidates. 
Particles with gravitational only interactions (newtorites) are another example. In this case the
sensitivity is quite poor and a short discussion is reported on possible improvements.

\end{abstract}


\end{frontmatter}

\thispagestyle{plain}

\section{Introduction}

\label{intro}
During the last decade a very large experimental and theoretical
effort has been devoted to understand the problem of dark matter
(DM). DM is necessary to explain the rotation of Galaxies and the measurements of the  velocities of
galaxies in clusters of galaxy.
The presence of DM is also predicted by the standard $\Lambda$CDM cosmological model that requires
a non relativistic $\Omega_{m}\sim$0.31 matter component   and 
a  baryonic component $\Omega_{b}\sim$0.05 \cite{Planck:2015xua}. Therefore the remaining fraction 
$\Omega_{DM} \sim $0.26 should consist of some
kind of weakly interacting matter different from the normal baryonic matter. The
local DM density in our galactic halo is expected to be of the order of
$0.3$ GeV/cm$^3$, and the DM speed should have values typical of
galactic halos, around $270$ km/s.

The leading dark matter candidates are supersymmetric thermal relics, a class of stable
weakly-interacting massive particles (WIMPs) that arise in supersymmetric theories; 
however searches for supersymmetry at  LHC and other accelerators have not found signals to date
\cite{Agashe:2014kda}. 
Likewise, direct detection experiments have yet to make any conclusive detection of conventional
WIMPs \cite{Akerib:2013tjd}.
The DAMA experiment has detected  a seasonal modulation in the signal that could be due to the dark
matter\cite{Bernabei:2013xsa}, but up to now 
no other experiment has confirmed this finding.
A possible signature of WIMPS could be the detection of $\gamma$ or neutrinos by WIMP - antiWIMP annihilation
inside celestial bodies. This search, called indirect detection, also has given no convincing
signature\cite{Agashe:2014kda}. Other DM candidates include
primordial black holes, monopoles, axions and sterile neutrinos.

Another scenario predicts that that DM particles could be much heavier than the few
TeV mass reached by current experiments. The possible existence of
super-heavy dark matter particles, sometimes called Wimpzillas,
would have interesting phenomenological consequences, including a
possible solution of the problem of cosmic rays observed above the
GZK cutoff \cite{GZK}. Many exotic names  are used in the literature for such
particles: Q-balls, mirror particles, CHArged Massive Particles,
(CHAMPs), self interacting dark matter, cryptons, super-weakly
interacting dark matter, brane world dark matter, heavy fourth
generation neutrinos,``MACROs", \emph{etc.} (see the references listed in
\cite{Bertone04}). Even if strongly interacting, these objects
could remain `dark' due to their large mass-to-surface area ratio
and correspondingly low number density required to explain the
observed DM mass density. There are some recent cosmological indications in favor
of strongly interacting dark matter  \cite{Rocha:2012jg,Peter:2012jh}.
However there is no experimental evidence supporting this idea.
Recent reviews are in Ref~ \cite{Jacobs:2014yca,Burdin:2014xma} and  \cite{Mack:2007xj}.

Composite objects consisting of light quarks in a
color superconducting phase have been suggested as super-heavy quark nuggets DM;  in addition,
super-heavy DM anti-quark nuggets could exist and could perhaps
solve the matter-antimatter asymmetry \cite{Gorham12}; the
detection of such anti-quark nuggets by cosmic ray experiments is
discussed in \cite{Lawson11}. The energy loss predicted for
super-heavy DM particles varies in different models, but it is
likely that for masses of the order of grams or more  such particles could be confused with meteors,
since the
velocity, $270$ km/s, is in the high-end tail of the meteor
velocity distribution\cite{Jem-Euso}.
Many names have been proposed for those objects: Quark nuggets, CUDOs \cite{Labun:2011wn},MACROs\cite{Jacobs:2015csa,Burdin:2014xma,Mack:2007xj}.

Here, we will focus our attention mainly on one possible kind of very
massive particle called ``nuclearite" \cite{derujula}. It is  an example of a well defined particle, consisting of neutral matter
with strange quarks among its constituents, and has 

 already been searched for by other experiment. Moreover the nuclearite results
can be easily extended to a generic strongly interacting particle.

Nuggets of Strange Quark Matter (SQM), composed of approximately the
same numbers of up, down and strange quarks could be the true ground
state of quantum chromodynamics \cite{Witten84}.
SQM nuggets could
be stable for all baryon numbers in the range between ordinary heavy
nuclei and neutron stars. They may have been produced in the early
Universe. 
SQM should have a relatively
small positive electric charge compared to that of heavy nuclei
\cite{derujula}. 
Macroscopic quark nuggets, neutralized by captured
electrons, are called nuclearites. 
Otherwise, as in the
case of small baryon numbers ($A \le 10^6$), assumed to be quasi
totally ionized, they will be called strangelets. There are several
concerns about the SQM hypothesis; one was raised in 1999, when
heavy-ion collisions between gold nuclei were produced at the
Brookhaven National Laboratory (USA) and, more recently, before the
LHC run with heavy ions: negative strangelets would
attract a positive nucleus and could swallow it, in a sequence that
could end with the digestion of the whole planet. Fortunately,
however, theoretical considerations suggest that
negative strangelets are unlikely to exist \cite{Dar99,Blaizot:2003bb}.

Finally it is important to note that  quite unlikely the DM would consist of only one particle with a
single mass and interaction; therefore, experimental searches should look at different
possibilities with a variety of techniques. 

        In this paper we will discuss particle detection using the thermo-acoustic effect (see below, sect.\ref{par:TAM}), and in particular with gravitational waves ($gw$) cryogenic bar detectors, expanding and improving the preliminary results presented in ref.\cite{Astone:2013bed}.
         We will focus on the Nautilus and Explorer detectors, that our group operated for decades.
We will then describe the analysis procedure and the selection criteria to identify candidates events, {i.e. those signal, among those recorded by the detectors as ``outliers", that are compatible with an excitation produced by one of the exotic particles described above.}
Then, starting from the energy distributions of the candidates events  we will present upper limits both 
specialized for nuclearites and for generic MACRO particles, vs. both mass and 
cross section, as suggested in \cite{Jacobs:2014yca}  and \cite{Jacobs:2015csa}. 

{In another scenario, DM only }interacts  gravitationally.
In this case the excitation of a bar detector is due directly to the newtonian force. The detection
mechanism is much more efficient, but signals are very {small} because the newtonian force is extremely
weak. The limits on this kind of DM with Nautilus and Explorer and the possible improvements will be
discussed in the last section. \\

\section { Gravitational wave bar detectors used as particle detector}

\subsection{Nautilus and Explorer}
\label{NaEx}
The   $gw$ bar  detector Nautilus\cite{nautilus} is located in Frascati (Italy) National
Laboratories of INFN;
it started  operations around 1998. {The current  run has been continuously ongoing since} 2003. The  detector
Explorer\cite{longterm},  similar to Nautilus,
was located in CERN (Geneva-CH) and was operational from 1991 to June 2010. 

Both detectors {work on  the same operating principles}.   Explorer and Nautilus consist of a large
aluminium alloy cylinder (3 m long, 0.6 m diameter) suspended in vacuum by a cable around its central
section 
and cooled to about 2 K by means of a superfluid helium bath.
The $gw$ excites the odd longitudinal modes of the cylindrical bar, producing an oscillation that is then mechanically amplified by
an auxiliary mechanical resonator tuned
to the same frequency which is bolted on one bar end face. This resonator is part of a capacitive
electro-mechanical transducer that produces an electrical a.c. current that is proportional to
 the displacement between the secondary resonator and the bar end face. Such current is then
amplified by means of a low-noise dcSQUID 
 superconductive device. 
 
 \begin{figure}[h]
\begin{center}
\includegraphics[width=12cm]{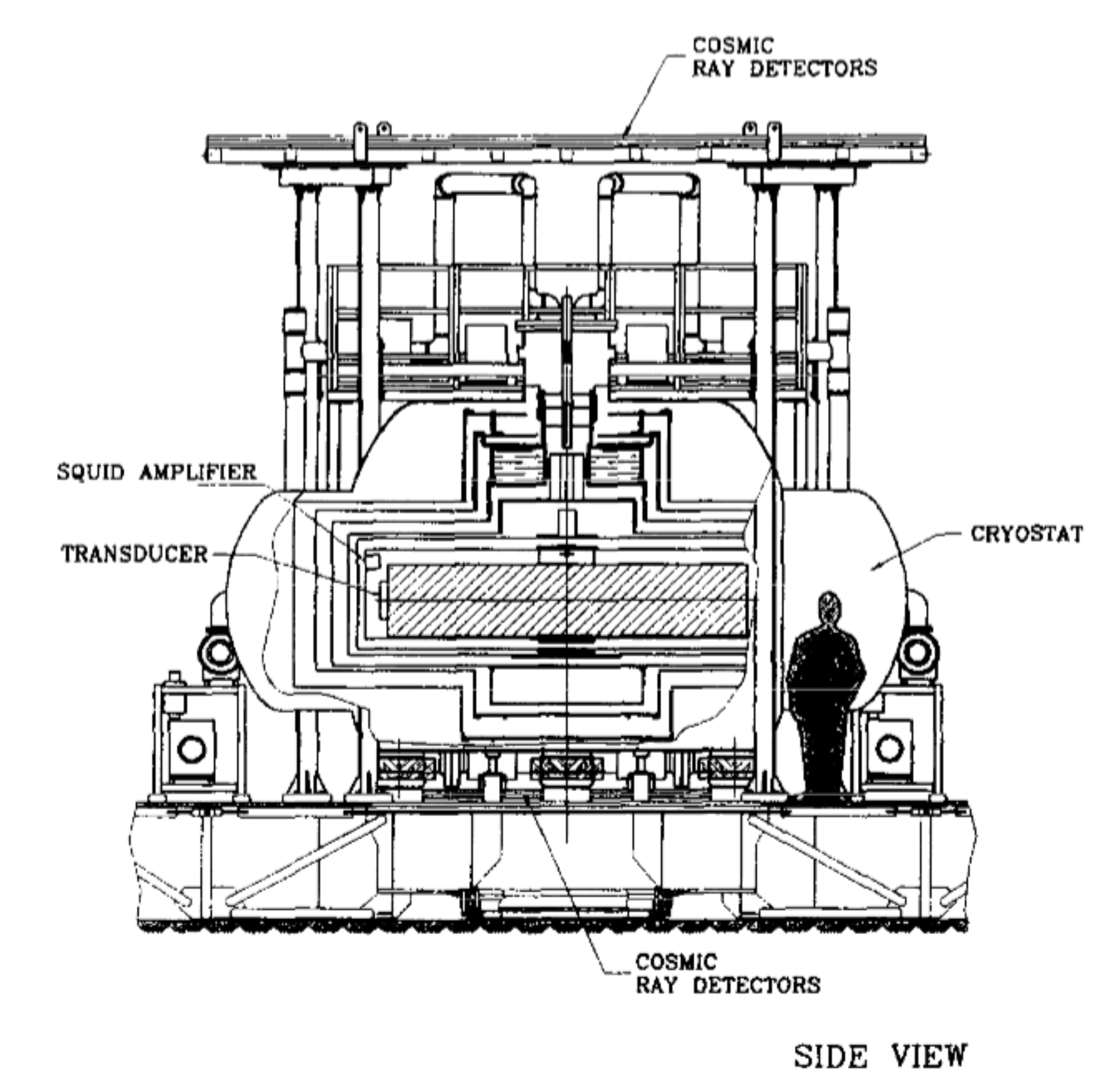}
\caption{ Schematic layout of the Nautilus $gw$ bar detector.}
\label{naut}
\end{center}
\end{figure}

 Central suspension and vacuum are used to reduce seismic and acoustic noise, while cooling to 2K reduces the thermal Nyquist noise of the electromechanical resonators.
Nautilus is also equipped with a dilution refrigerator that enables operations at 0.1 K, further
reducing the thermal noise. 
However,  after 2001  the refrigerator {was kept idle, because its operation negatively affected the detector duty cycle. }

The output of the SQUID amplifier is conditioned by band pass filtering and by an anti-aliasing
low-pass filter, then sampled at 5 kHz and stored on disk. 
 Sampling is triggered by a GPS-disciplined rubidium oscillator, also providing the time stamp for
the acquired data.
The data are processed off-line, applying adaptive  frequency domain filters optimized for delta-like signals, { i.e. very short (few ms) bursts}.
 We first whiten the data, i.e. remove the effect of the detector transfer function.  A filter
matched to delta 
excitations is then applied to this stream. The filter is designed
and optimized for delta-like signals, but it works equally well  for a wider class of short bursts,
like e.g. damped sinusoids with decay time less than 5 ms.  
 The noise characteristics estimate is updated averaging the output over 10 minutes periods.
Traditionally, {in the antenna jargon, the noise energy is expressed in kelvin units (1~K~=~$1.38 \times 10^{-23}$~J)}.  The typical {filtered }noise of  data
considered in this paper is between 1 and 5 mK.

\subsection{The Thermo-Acoustic Model}\label{par:TAM}
The interaction of energetic charged particles with a normal mode of an extended elastic cylinder
has been extensively studied over the years, both on the theoretical and on the experimental
side\cite{Nim2011}.

According to the so called  ``thermo acoustic model"  a particle crossing the bar produces  longitudinal
vibrations  originated from the local thermal expansion due to the warming up due to the energy loss.
In particular, the vibration amplitude {linearly depends on} 
the ratio of two thermo-physical
parameters of the material,  namely the thermal expansion coefficient and the specific heat at
constant volume. The ratio of these two quantities appears in the definition of  the Gr\"{u}neisen
parameter $\gamma$.
 It turns out that while the two thermo-physical parameters vary with temperature, $\gamma$
practically does not, provided the temperature is above the material superconducting $(s)$ state 
critical temperature.  

{The change in vibrational energy  $\Delta E$} of the fundamental vibrational mode due the energy loss of a particle
crossing an aluminium cylindrical bar is \cite{allega,deru,liu}: 

\begin{equation}
 \Delta E=\frac{4}{9\pi}\frac{\gamma^2}{\rho L v^2}(\frac{dW}{dx})^2 
 \times [sin(\frac{\pi z_o}{L})\frac{sin[(\pi l_ocos(\theta_o)/2L)]}{\pi Rcos(\theta_o)/L}]^2
 \label{eliub}
\end{equation}

where  $L$ is the bar length, $R$ the bar radius, $l_o$ the length of the particle track inside the
bar, $z_o$ the distance of the track mid point  from one end of the bar, $\theta_o$ the angle
between the particle track and the axis of the bar, $\frac{dW}{dx}$ the energy loss of the particle
in the bar, $\rho$ the density, $v$ the  longitudinal  sound velocity in the material.
This relation  is valid for the  normal-conducting $(n)$ state material and some authors (see the
references in  \cite{Nim2011}) have extended the model to a super-conducting  ($s$) resonator,
according to a scenario in which the vibration amplitude is due to two pressure sources, one due to
$s-n$ transitions in small regions centered around the interacting particle tracks and the other due
to thermal effects in these regions now in the $n$ state.

 It is important to note, at this point, that  a $gw$ bar antenna, used as particle detector, has
characteristics very different from the usual particle detectors which are sensitive to ionization
losses:  indeed an acoustic resonator can be seen as a zero threshold calorimeter, sensitive to a
vast range of energy loss processes. The usual $gw$ software filter works well up to a time scale of
the order of 5 ms, corresponding to a $\beta {=v/c} = 4 \times 10^{-6}$ for a 60 cm particle
track. So the antenna is sensitive  to very slow tracks: this is another very important difference
with respect to the usual particle detectors.

Due to this effect cosmic ray showers can excite sudden mechanical vibrations in a metallic cylinder
at its resonance frequencies; in experiments searching for  $gw$ these disturbances are hardly
distinguishable from the searched signal and represent an undesired source of accidental events,
thus increasing the background. This effect was suggested many years ago and a first search was
carried out with limited sensitivity {on} room temperature Weber type resonant bar detectors and
ended  with a null result~\cite{Ezrow:1970yg}. 
{To monitor these signals and to provide effective vetos, Nautilus was equipped with a cosmic ray shower telescope system, composed of streamer tubes detectors positioned both above and below the cryostat. For further details see ref.\cite{cosmico3}.
}
 The first detection  of cosmic ray signals in a $gw$ antenna took place in 1998, with the  Nautilus detector
operating at a temperature $T$~=~0.14~K~\cite{cosmico1},  i.e. below the superconducting ($s$)
transition critical temperature $T_c \simeq$~0.9~K.  During this run, {an unexpectedly large number of events with} very large
amplitude were detected.  This result suggested an anomaly either in the thermo-acoustic model or in the cosmic ray
interactions\cite{cosmico2}. However the observation was not confirmed in the 2001 run with Nautilus
at $T=1.5$ K~\cite{cosmico3} and therefore we {formulated} the hypothesis that the unexpected behavior was
due to the superconducting state of the material. 

 This result  prompted the construction, in 2002, of a scintillator cosmic ray  detector  also
for the Explorer $gw$ detector \cite{Astone:2008xa} as well as the beginning of a dedicated experiment
(RAP)~\cite{rap,rap1,rap2}, that was carried out at the INFN Frascati National Laboratory to study the
vibration amplitude, caused by the hits  of a 510 MeV electron beam, in a small Al5056 bar. The
experiment was also motivated by the need of a better {knowledge of the  low temperature values of the thermo-physical parameters
for the alloy Al5056, used in the bar detectors. }

\begin{figure}
 \begin{center}
  \includegraphics[height=60mm,width=80mm]{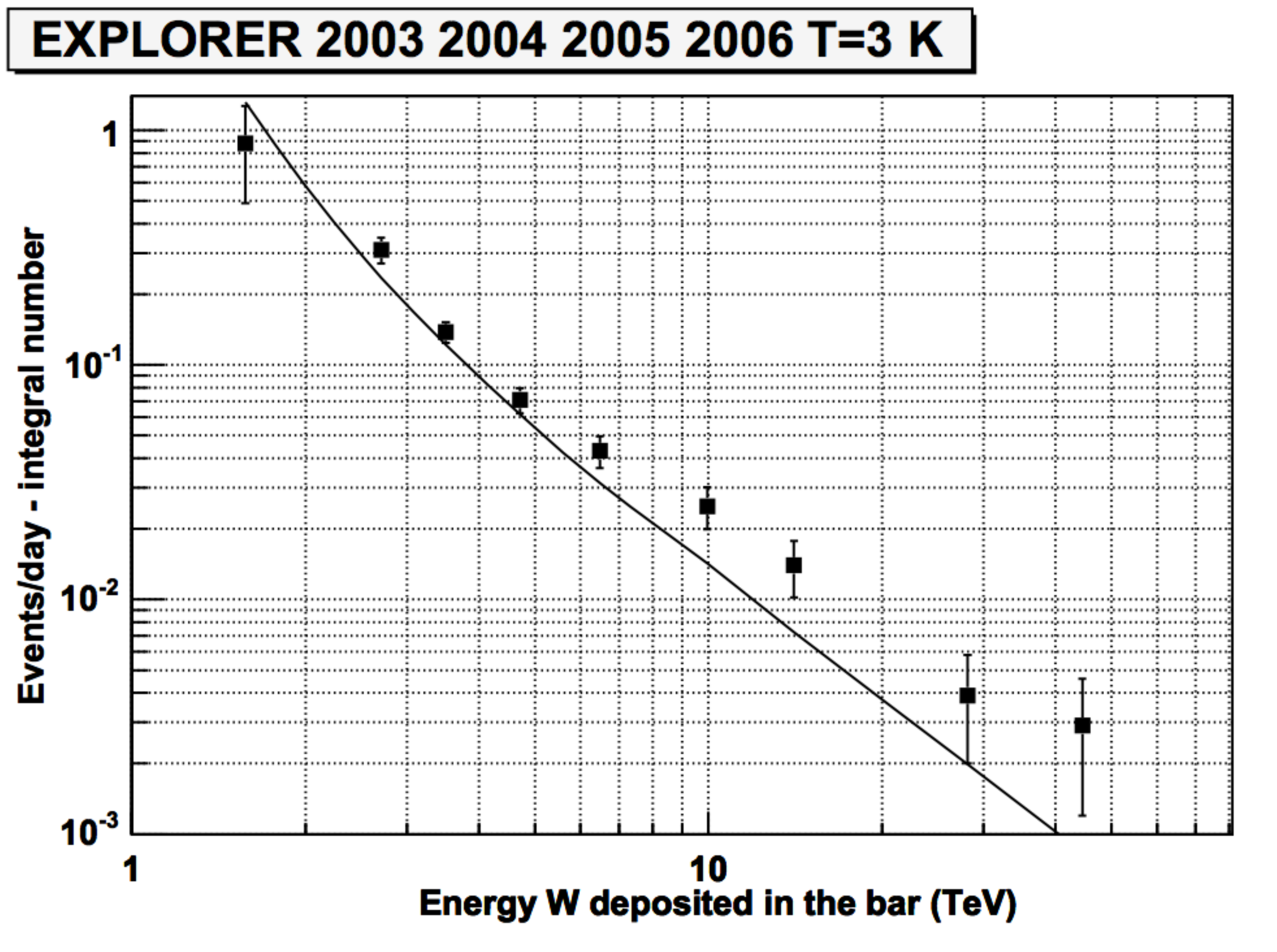}
  \end{center}
  \vspace{-0.5cm}
  \caption{Integral distribution of extensive air showers in Explorer. The line shows the prediction
based on the thermo acoustic model \cite{Astone:2008xa}. One  event, the largest ever recorded, with an energy of 360 TeV. ( 670 K in
kelvin units), sits at the far right outside this plot.}
  \label{fig:shower}
\end{figure}

{ As a result of these experiments we have a  model capable to predict the signal produced in a bar like that of Nautilus or Explorer by the interaction with a passing energetic particle:

 for a particle that  loses a total amount of energy $W$ while intersecting the bar, orthogonally to its axis and through its middle section,
 the energy of the longitudinal fundamental mode of vibration will change by :}

\begin{equation}
\Delta E \sim7.64\times10^{-9}~W^2~\delta_G^2 \hskip1cm  [\rm{K}]
\label{ww}
\end{equation}

\noindent { where we assume negligible the bar oscillation energy before the interaction and
we express  $\Delta E$  in
kelvin units}  and $W$ in GeV units. The numerical constant  has the value computed using
the Gr\"{u}neisen parameter $\gamma$  of pure aluminium at 4~K.  {The parameter $\delta_G$
describes the correction due to different temperatures or materials:}

the measurements, carried out on Al5056 alloy \cite{Nim2011} , give:
$\delta_{G,n}$~=~1.16  for T=4K and $\delta_{G,n}$~=~1.3   for $T\sim$1.5 K above the  transition
temperature  ( $T_c = 0.845 K$), with a relative uncertainty of 6\%.  For superconductive Al5056  we found $\delta_{G,s}~=~5.7 \pm 0.9$.
Most of the events
presented here {were gathered} with the antenna in equilibrium with a bath at   $T\sim$1.5-2  K.
 
 Using the results of these measurements  we have a good agreement both in rate and amplitude of the
extensive ray shower detected in Nautilus and Explorer 
\cite{Astone:2008xa} as shown in fig.\ref{fig:shower},  with the expectation based on  cosmic ray
physics and the thermo-acoustic model. This reinforces our confidence in 
a full understanding
 of the $gw$ bar detectors used as particle detector.
  
\section
{Dark Matter signals in gravitational wave bars}

\subsection{ MACROs and nuclearites}

The interaction with the bar of massive candidates with radius  $R$ much larger than any microscopic length
scale, e.g. the electron's Compton wavelength or the Bohr radius, is quite simple 
because we can  ignore any quantum-mechanical aspects of scattering, and any short-range
interaction: the interaction cross-section $\sigma$ corresponds then to  the dark matter
geometric cross-section.
In ref \cite{Jacobs:2014yca} the name MACROs was proposed for  this class of macroscopic dark matter.
If the MACRO candidate carries a net charge,  it may also interact electromagnetically. In this section we
consider only neutral matter.  Additional contribution to the energy lost by charged matter or
antimatter will increase the expected signal.

The main energy loss mechanism for a generic neutral  particle,  having  
cross section  $\sigma$,  a non-relativistic velocity $v$ and mass $M\gg M_{atomic}$, is by atomic collision:

\begin{equation}
\label{eq:rel}
\frac{dE}{dx}= -\sigma \rho v^{2}
\end{equation}

\noindent where $\rho$ is the density of the traversed medium and $v$ the particle velocity. In pure
aluminium  $\rho$ =2700 $kg/m^3$.
In the case of a compact object  the cross section $\sigma$ is the effective area $A$ than can be
computed via the  compact object density
$\rho_{N}$.

When a MACRO crosses the atmosphere (or the Earth), it loses energy and the velocity decreases
according to:

\begin{equation}
v= v_{0}e^{-\frac {\rho_a L \sigma}{M}} 
 \label{eq:vel}
\end{equation}

where $\rho_a$ is the atmosphere density, L is the path {length}, $\sigma$ and $M$ {the particle} cross section and mass. In order to be
detectable in an analysis where   {delta} like signals are selected,  these excitations should be in the ms range
and therefore the velocity $v$ should be $\gtrsim$ 1 km/s.

Consider  a  MACRO  of velocity $c\beta$, intersecting orthogonally  through the center of the Nautilus
(or Explorer) bar: by applying  eq.\ref{ww} we can compute the energy change $ \Delta E$ of the bar fundamental mode (the ``signal"):

\begin{equation}
\label{eq:nume}
\Delta E= 1.08  \cdot 10^{32}(   \frac{\beta} {10^{-3}} )^4  \sigma^2  \hskip1cm  [\rm{K}]
\label{desigma}
\end{equation}
Here $\sigma$ is in $cm^2$ and we have used $\delta \sim 1.3$ at $T =2K$   \cite{Nim2011},
 and $ \Delta E$ is measured in kelvin.  So,  an object having atomic size, as a nuclearite, produces
signals in the 1 K region. 
Nuclearites with galactic velocities are protected by their
surrounding electrons against direct interactions with the atoms
they might hit.
 For a small nuclearite of mass less than $1.5$ $ng$, the
cross-section area $A$ is controlled by its electronic atmosphere whose radius is never smaller than
$10^{-8}$ cm:

\begin{equation}
\label{eq:sigma}
A =\left\{ \begin{array}{ll}
\pi \cdot {10}^{-16} \,\textrm{cm$^{2}$}    &  for ~ M < 1.5 \, ng
  \\
\pi {\left(\displaystyle{\frac{3 M}{4 \pi {\rho}_{N}} }\right)}^{2/3}  \,\textrm{cm$^{2} $} & for ~ M > 1.5 \, ng
\end{array} \right.
\end{equation}
where \mbox{$\rho_{N}= 3.6\cdot10^{14}\,g/cm^{3}$} is the nuclearite
density and $M$ its mass \citep{derujula}.

Consider indeed  a  nuclearite  of mass $M$ and velocity $c\beta$, intersecting orthogonally the center of
the Nautilus (or Explorer) bar; by  applying eq.\ref{desigma} we have:
\begin{equation}
\label{eq:nume}
\Delta E= 10.7 (   \frac{\beta \theta(M)} {10^{-3}})^4  \hskip1cm  [\rm{K}]
\end{equation}
where  $ \Delta E$ is the energy variation of the bar fundamental mode measured in kelvin  and
$\theta(M)=(M/1.5~ng)^{1/3}$ if $M>1.5$ ng. Otherwise $\theta(M)=1$. 

The location of the detector used has impact on the range of $\sigma$ and $M$ that can be detected.
For example according to eq.\ref{eq:rel}, nuclearites having galactic velocity (about 300 km/s)
and mass heavier than $10^{-14}$ $g$ penetrate the atmosphere, while
those heavier than 0.1 $g$ pass freely though an Earth diameter.

\subsection{Newtorites}
Particles having newtonian interaction only can be called newtorites.
Due to the long range nature of the newtonian force, signals could occur even if the particle
does not 
cross the bar. In the case of a  point like particle moving with a constant velocity $v$ along a
straight trajectory coming from infinity and going to infinity,  the vibrational amplitude of the
nth-vibrational mode is given by\cite{deru}:

\begin{equation}
A_{n}= -\frac {2GM}{Vv} \int_{V}   \frac { \mathbf {u_{n} \cdot }  
  \mathbf {(x_{T}- x_{T}^{0}) } }  { (x_{T}- x_{T}^{0}) ^2} d^3\mathbf{x}
\label{eq:Anewton}
\end{equation}
Here $G$
is the gravitational constant, $M$ the mass of the newtorite,
$\bf{x_{T}}$  are the transverse coordinates of a volume element of the detector relative to a fixed point $\bf{x_{T}^{0}}$, arbitrarily chosen along the particle track ; $\bf{u_n}$    is the spatial part of the nth
normal-mode oscillation normalized to the volume $V$ of the bar.
For a thin bar with radius $r$ and length $L$,  ($r \ll L$)  
$\bf{u_n}$   can be approximately written, using cylindrical coordinates \cite{liu}:

\begin{eqnarray}
\label{eq:un}
& u^r_n= \sqrt 2 \sigma_{P} \pi (r/L) sin(n\pi z/L) \nonumber \\
& u^z_n= \sqrt2  cos(n\pi z/L)
\end{eqnarray}
Here $\sigma_{P}$ is the aluminium Poisson module.  The energy variation  in the bar is obtained by: 
\begin{equation}
\label{eq:Tnewton}
 \Delta E_{n}= \frac {1} {2 k_B } \rho A_{n}^2 V  \hskip1cm  [\rm{K}]
\end{equation}
Here {$k_B$} is the Boltzman constant.
In this paper we are only interested  in the first longitudinal mode n=1, and we assume
that the velocity $v$ of the particle is large enough  that most  of the signal is contained in a few ms.
This requirement is due to the $\delta$-like filter used to extract the antenna events.
(different filters could in principle detect longer signals,  lasting  up $\sim40 s$).
The signal is a fairly complicated function of the newtorite's 
trajectory and has been computed in \cite{deru} in the particular case of orthogonal trajectory in the
middle of the bar,  and  for $r/L\to0$. In this case we can put  $u^r_1=0$ and we obtain:
\begin{equation}
\label{eq:Tmiddle}
 \Delta E \thicksim 30  \pi r^2 \frac{ \rho G^2 } {k_B  L} (\frac{M}{v})^2   \hskip1cm  [\rm{K}]
\end{equation}
Numerically we have for  Nautilus  $ \Delta E \thicksim  2.4 (M/v)^2$ K with $M$ expressed in kg and $v$ in
km/s.
 
\begin{figure}
 \begin{center}
  \includegraphics[height=60mm,width=80mm]{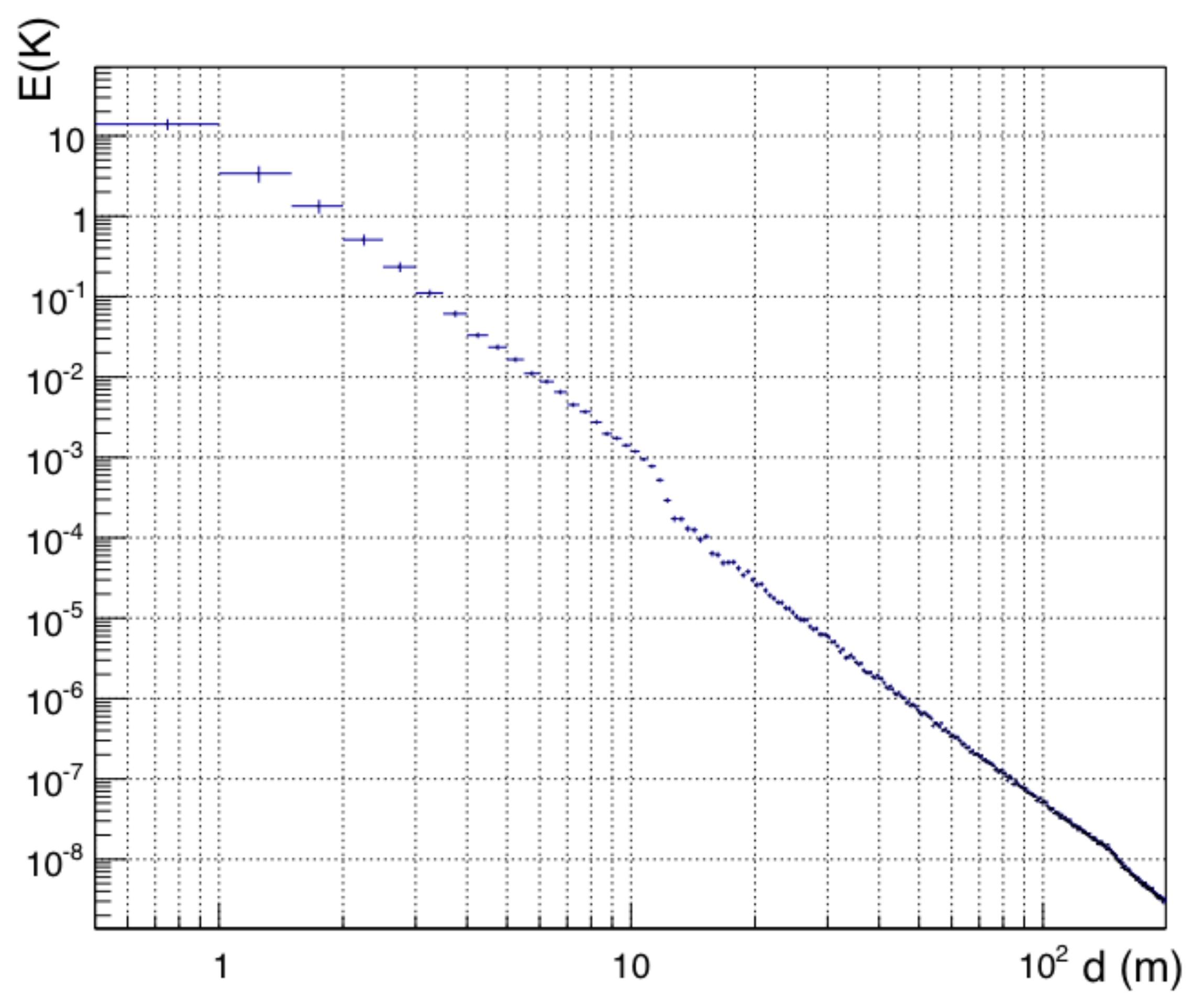}
  \end{center}
  \vspace{-0.5cm}
  \caption{Average from different directions of the signals due to a $M/v=10kg~s/ km $ newtorite vs. distance from the bar center}
  \label{fig:Edist}
\end{figure}

In the general case the signal has been computed by numerical integration of eq.\ref{eq:Anewton}
inserted in a Montecarlo to simulate random directions. The result of one of those calculations as
function of the distance of the trajectory from the bar center and for $M/v$=10  kg~s/km is shown in
Fig~\ref{fig:Edist}. From this figure we can see that, at large distance d, the signal
energy scales as 1/d$^4$,  as expected from eq.\ref{eq:Anewton}. The signal falls below the energy threshold used in this analysis (see the next paragraphs) for d larger
than $\sim$ 3 m for $M/v$=10 kg~s/km. For $M/v$=100 kg~s/km this threshold occurs at $d \gtrsim$ 10 m.
 
\section{Data processing and selection}
\label{proc}

{In this section we describe the procedure to select  large signals that emerge, in a statistical sense, from the noise of the antenna. 
We shall call them candidate events or simply events, but this does not imply any assumption on their origin or cause.
{As mentioned in sect.\ref{NaEx}, for this
analysis we considered the stream of delta-filtered data sampled at
5 kHz.}
Indeed, the signal expected from the interaction of a nuclearite with the bar has
the characteristics of the response to a delta-like excitations, and we  take
advantage of this to reject events not compatible with a delta. In this respect, the search for 
this class of events is more demanding than the usual coincidence searches for g.w. candidates.

In the filtered
data we search for {\it candidate events} by applying a
threshold, usually set at 4 times the RMS value, continuously
updated in a exponentially weighted manner. This produces a large
number of events, a few thousands per day, with a gaussian amplitude
{distribution}, but with a tail of high amplitude events consistently
higher than for a gaussian. In the past, this behavior has been
extensively studied and a number of {possible causes} for this instrumental
extra-noise was identified: seismic excitations, instabilities of
the SQUID electronics, electrically induced spikes, shaking due to
the cryogenic fluids, cosmic rays showers, and so on.

In this work, we have considered only the events with energy
larger than $0.1~K$. As the RMS noise is of the order of a few mK,
the number of these large signals reduces from thousands to tens
 per day.

The data processing applied for the present work can be divided in
two parts: first we select the periods of ``good" operation
and then we proceed with the identification of {spurious events}

\begin{figure}[h]
\includegraphics[width=7cm,height=8cm]{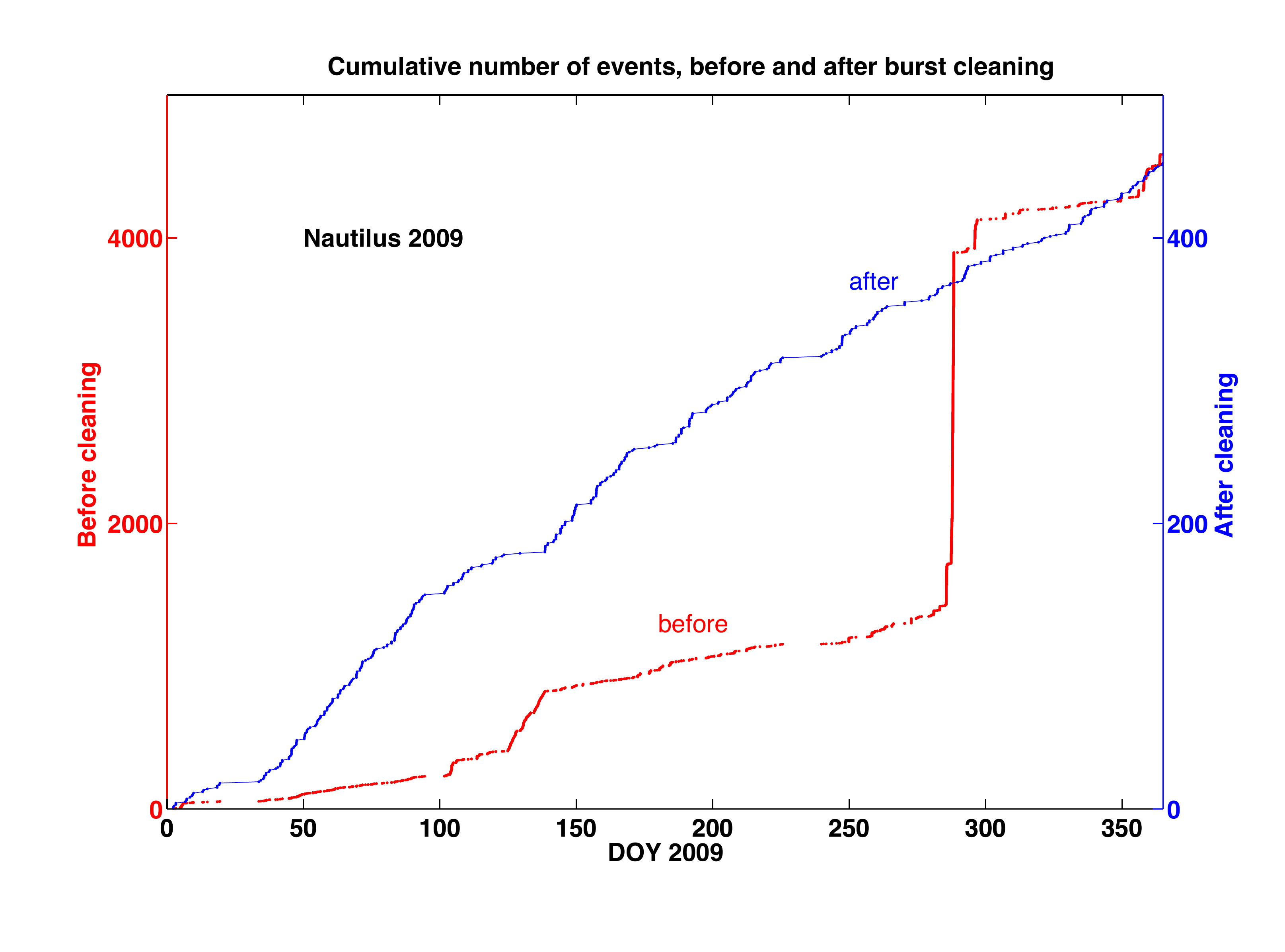}
\includegraphics[width=7cm,height=8cm]{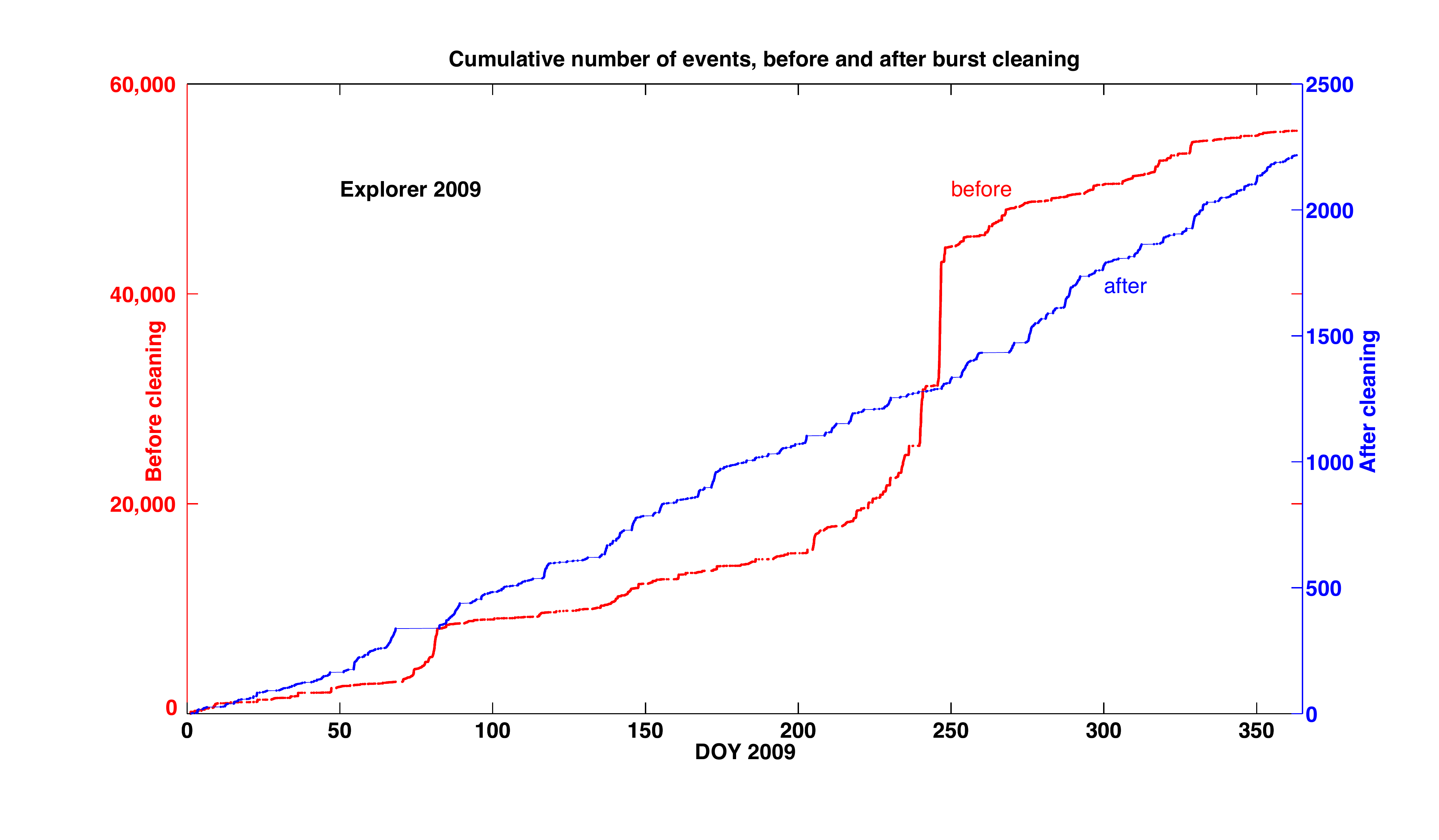}
\vspace{-0.5cm}
\caption{  Cumulative time distribution of the
Explorer and Nautilus events during year 2009. One can clearly see that there
are periods with a rate of events orders of magnitude larger
than in the quiet periods of operation.
For each detector, we show the distribution before (red online) and after (blue online) the cuts operated to remove the noisy periods.
The selections reduces the number of events by a factor between 10 and 20.
}
\label{fig:cday}
\end{figure}

\subsection{Selection of periods of operation}
\label{proc1}

The first selection step  is to remove the periods that were marked by
operator's flags or comments, indicating that maintenance operations were going
on. Further cuts are performed by looking at the amplitude, in the unfiltered
data, at the frequency of a {reference tone} that is added in the
SQUID electronic for the purpose of monitoring the gain of the whole
electronic chain. These two cuts
usually reduces the livetime by $\simeq 20-30 \%$

A {third} selection is based on the seismic noise as detected by some
accelerometeres mounted on the bars cryostat and recorded by an
ADC sampled at 0.1024 Hz. When the seismic noise exceeds a previously
set threshold, a period of at least 10 s is vetoed. This cut
removes a negligible amount of livetime, but several large events.

A fourth selection considers  the value of the 
{\it effective temperature} $T_{eff}$, 
i.e.,  the variance of the filtered data, measured in $K$.
As previously said, this value was usually around 1-2 mK for Nautilus
and 2-3 mK for Explorer. We 

{ vetoed} periods of at least 10
minutes when  $T_{eff}$  continuously exceeded a given threshold,
usually 2.5 mK for Nautilus and 5 mK for Explorer. This cuts only a
$1-2 \%$ of the livetime.

A fifth selection is done observing the time distribution of the candidate
events, that shows the presence of bursts of events, having duration
between 1 or few days down to few minutes (see fig.\ref{fig:cday}).
This behaviour is clearly due to some malfunctioning of the detector,
not otherwise characterized.
The technique employed is to consider periods of time of different
duration (usually 1 day, half day, 1 hour, 10 minutes) spanning
continuously the livetime, and to count the number of events in each period.
We then veto the periods where the number of events found is
much higher than that predicted by a Poisson distribution based on
the average rate, usually giving a Poisson probability of $10^{-5}$
or less. This selection is the one resulting in the larger reduction
in the number of events, between 50 and 90 $\%$, while reducing
the livetime by $10-30 \%$.

A last selection requires an uninterrupted operation for a given
period, lasting in most cases at least 10 hours. In some cases,
this requirement was relaxed down to 5 or 3 hours, in order not
to cut more than $30 \%$ of the livetime.

{The overall cuts produced by these conditions range
between 10$\%$ and, in few very noisy periods, 70$\%$ of the livetime, and between 96 and 99.8 $\%$ of the number of events.
}
\subsection{ 
Event selection - Rejection of false candidates events}
\label{proc2}
\begin{figure}
\centering
\begin{tabular}{cc}
\includegraphics[width=12cm]{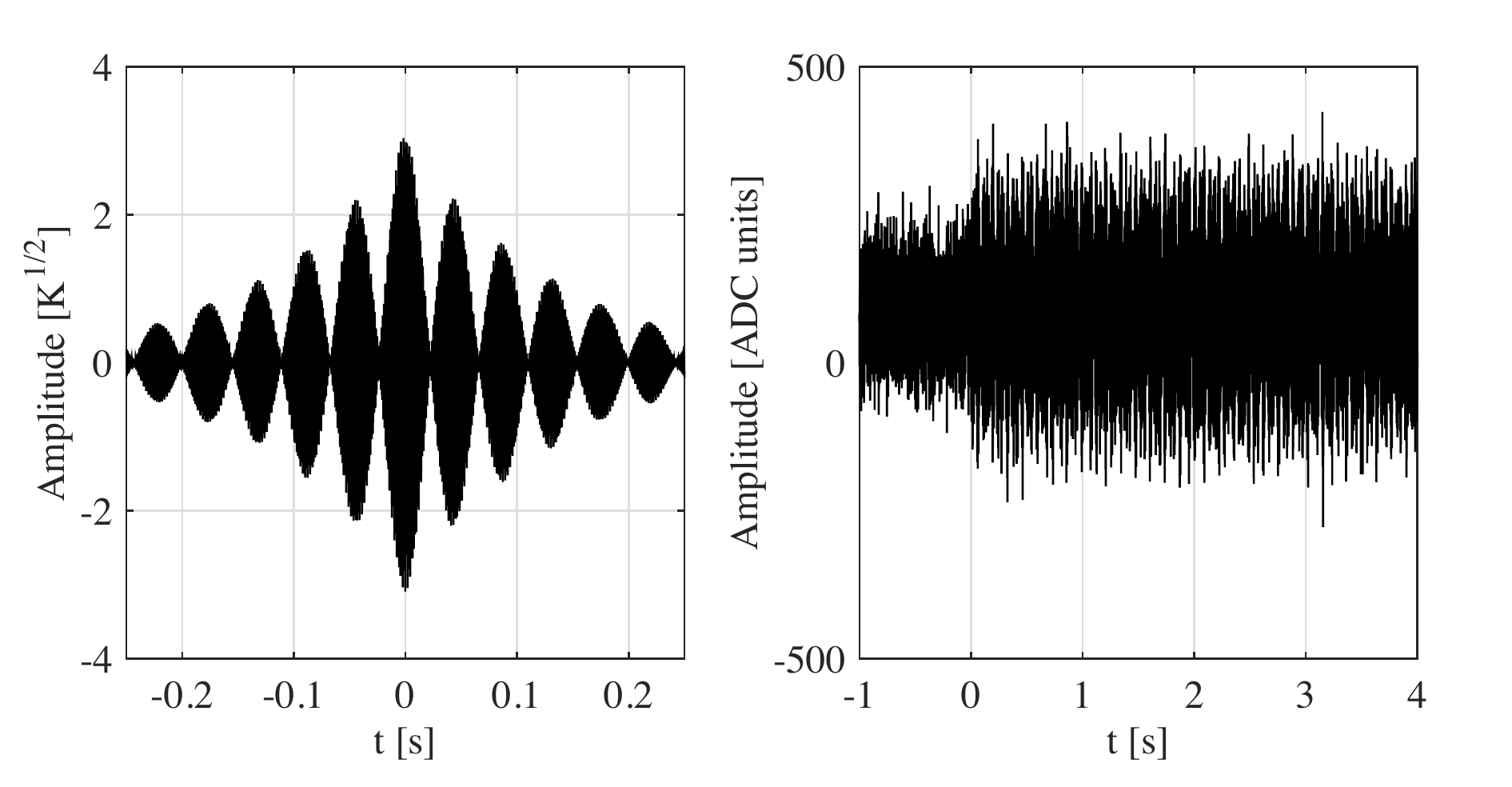}&
\end{tabular}
\caption{An Explorer event due to a cosmic ray shower: filtered
output (left) and raw ADC data (right).}
\label{fig:ex_cosm}
\end{figure}

\begin{figure}[h!]
\centering
\begin{tabular}{cc}
\includegraphics[width=12cm,height=6cm]{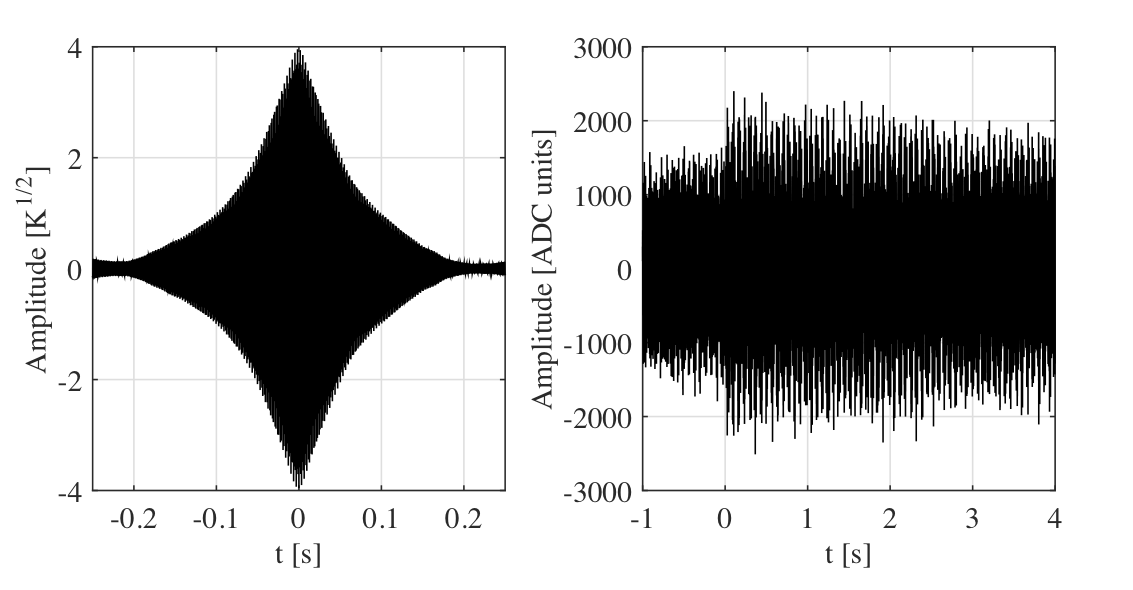}
\end{tabular}
\caption{A Nautilus event due to a cosmic ray shower: filtered
output (left) and raw ADC data (right).}
\label{fig:na_cosm}
\end{figure}

The relatively small number of events, together with their large
amplitude, allows us to analyze each single event and to
select those compatible with the assumption to be produced by a very
short excitation, such as we expect to be caused by a nuclearite.
Figures \ref{fig:ex_cosm} and \ref{fig:na_cosm} show, for Explorer
and Nautilus, what a real delta-like event should look like. In both
cases, {they are real events, produced by a large cosmic rays
shower.  Note the different shape of the filtered response in the two detectors:  
it depends on the details of the transfer functions of the two coupled mechanical oscillators (bar and auxiliary resonator) as described in sect \ref{NaEx}.} 

A first selection rejects the events vetoed by a coincidence with signals
from the cosmic rays detectors. In order to identify an event as
due to {a cosmic ray shower}, and therefore to exclude it from the possible nuclearite
events, we require a large number of particles present in both
the particle detectors (over and underneath the bar) and a strict
time coincidence (less than 10 ms) between their signals and
the bar event. This selection takes out only a few events per year.

For each event, we analyzed both the delta-filtered data and
the ``raw" data, that is the unfiltered ADC readings.

Examining the filtered data, we reject an event if its shape is judged 
really incompatible with a delta excitation plus
noise. 
Fig.\ref{fig:bad1} shows some examples of events rejected for this reason.

{Candidate events that do appear, after filtering, as delta-like signals, can
also be rejected by looking at the direct ADC data: indeed,}
 we saw a large number of strange
behaviors that can produce the appearance of a delta-filtered event.
 There are cases where a very short
(few samples) peak appears in the ADC data, most likely due to an
electrical spike either in the main power or induced on the circuitry
by some electromagnetic noise (lightnings, for example). 
In order to quickly identify  these instances, we carried out an event
search in the ADC data, requiring
the presence of a small number of samples above {an adaptive} threshold.

Then, we perform the coincidence search
between these events and the filtered ones, rejecting those in strict
coincidence (50 ms).  

Another case appears sometimes with the ADC reading dropping  to about zero
for {a few seconds}: we know that  {  this happens when the SQUID ``unlocks", i.e. suddenly changes its  operating point;
indeed we had a veto monitor based on the reading of the SQUID working point, that }
we used to remove short periods around the times when these jumps occur.

In other cases, but this is apparent only for very large events,
we see that the ADC amplitude decays with a time constant much shorter
than the one associated with a real mechanical excitation. This
 happens when an instability causes a resonance in the electrical
circuit of the bar transducer.
Fig.\ref{fig:bad2} shows examples of this class of rejected events.

Finally, there were few cases when the ADC showed a really erratic
or unstable behaviour.

\begin{figure}
 \begin{center}
  \includegraphics[height=70mm,width=120mm]{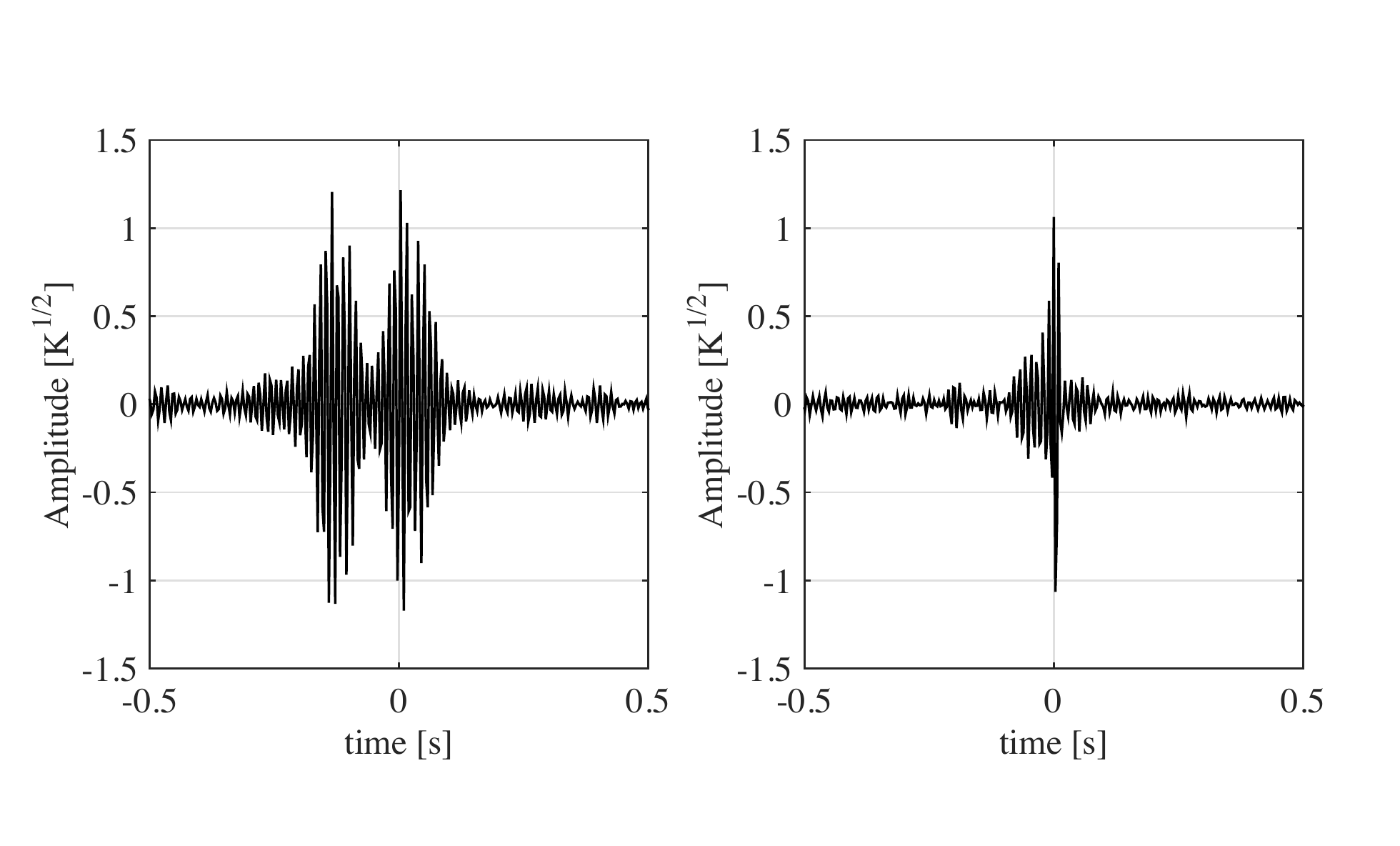}
  \end{center}
  \vspace{-0.5cm}
  \caption{Two examples of filtered events that were rejected due to their shape, bearing no resemblance to the response of Nautilus to a delta signal}
  \label{fig:bad1}
\end{figure}

\begin{figure}
 \begin{center}
  \includegraphics[height=70mm,width=120mm] {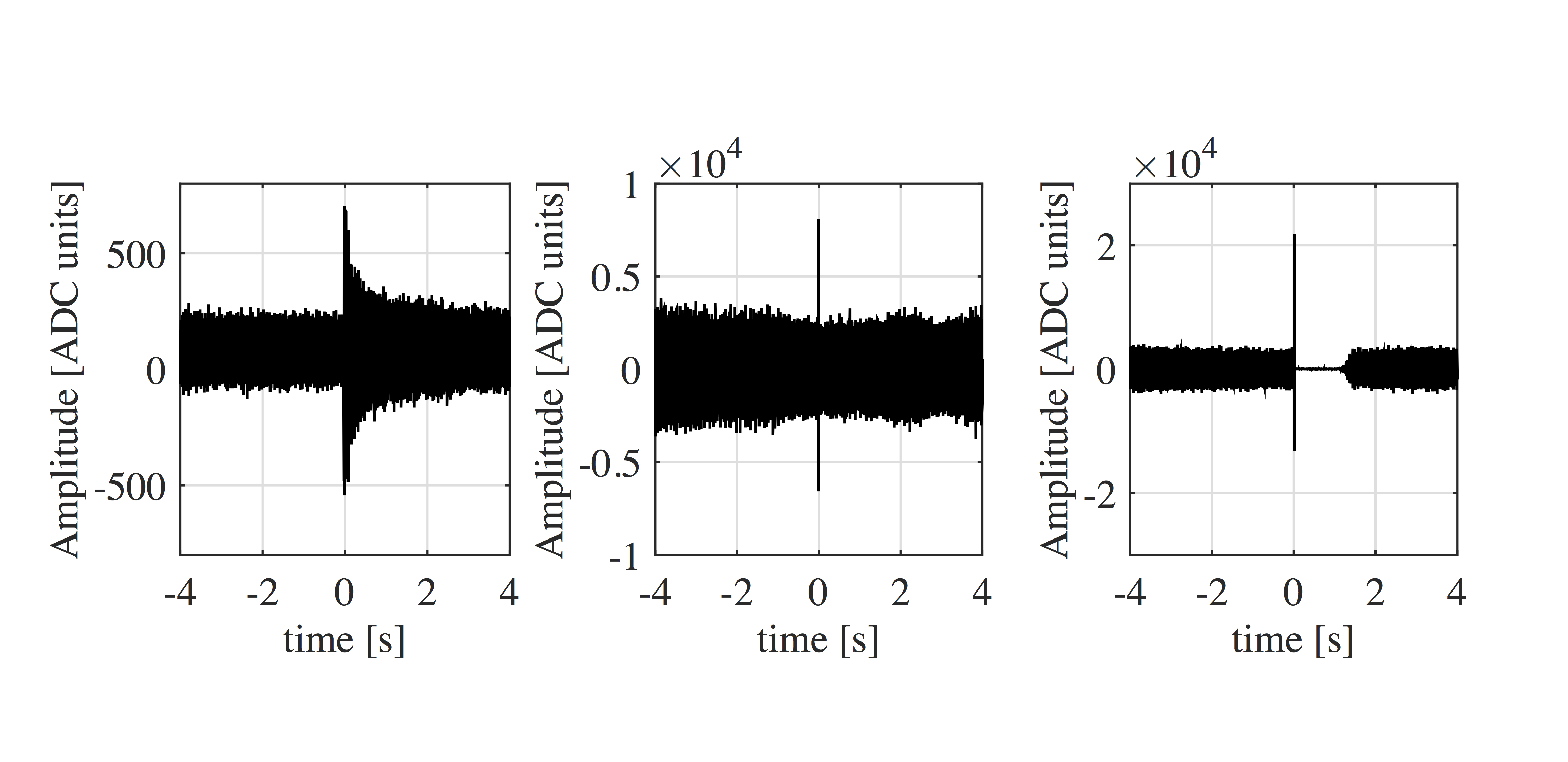}
  \end{center}
  \vspace{-10mm}
  \caption{{Three examples of  events rejected by inspecting the ADC, unfiltered output. From left to right:  a decay too short in Explorer (left), a spike lasting few samples, in Nautilus (center), the output going to zero, in Nautilus (right)}}
  \label{fig:bad2}
\end{figure}

{The final result is that these criteria allowed us to reject, as ``unacceptable events", a
fraction starting from 7$\%$ and  up to 93$\%$ (depending on detector and period of observation) of the events in the previously  defined periods of good operation.}

\subsection{Results on the rate of events}
\label{proc3}

For this work, we have considered the Nautilus data from 2005
to 2014 and the Explorer data from 2005 to  the end of its operations, in June 2010.
Tab.\ref{tab_years} summarizes the results of the present analysis.

As mentioned in sect.\ref{proc1}, the Explorer data were in general of poorer
quality than the Nautilus ones: not only the average
$T_{eff}$ is higher, but the high amplitudes tail is larger and
there are more instabilities and interruptions. This can be easily
seen looking at the results shown in Table\ref{tab_years} and it is the
reason why we consider the two detectors as ``different" and decided not to
merge their respective results.

Besides, each detector had varying performances over the
full periods in analysis, as shown in figs.\ref{fig:rate} where we plot the
{cumulative} rates of events vs their energy for the two detectors and
for the different years of analysis.

\begin{table}
\centering
\begin{tabular}{||c||c|c|c||c|c|c||}
\hline\hline
 & \multicolumn{3}{c||}{\bf Explorer} &  \multicolumn{3}{c||}{\bf Nautilus} \\
\hline\hline
 & & &  & & & \\[-3mm]
 Year & Livetime (d) & N.ev. & Rate (ev/y) & Livetime (d) & N.ev. & Rate (ev/y) \\
\hline\hline
 & & &  & & & \\[-3mm]
 2005 & 153.65870 &  78  &  185.41 & 186.6178  &  84  & 164.41 \\
 2006 & 126.03640 & 107  &  310.08 & 135.9331  & 107  & 287.51 \\
 2007 & 108.28680 &  57  &  192.26 & 152.9947  &  67  & 159.95 \\
 2008 & 131.68350 & 109  &  302.33 &  93.0710  &  79  & 310.03 \\
 2009 & 100.70430 & 167  &  605.70 & 191.6569  & 103  & 196.29 \\
 2010 &  38.18758 & 110  & 1052.11 & 206.9278  & 100  & 176.51 \\
 2011 &           &      &         & 287.8376  & 160  & 203.03 \\
 2012 &           &      &         & 300.6588  & 121  & 146.99 \\
 2013 &           &      &         & 289.9617  & 110  & 138.56 \\
 2014 &           &      &         & 283.4931  & 286  & 368.48 \\
\hline\hline
 Tot  & 658.55728 & 628  &  348.31 & 2129.1525 & 1217 & 208.77 \\
      & (1.803 y) &      &   (avg) & (5.829 y) &      &  (avg) \\
\hline\hline
\end{tabular}
\vskip 0.1 in
\caption{Characteristics of the bulk of data analyzed, with the final
livetimes, number of events and rates.}
\label{tab_years}
\end{table}

\subsection{False dismissal and efficiency}
\label{proc4}

\begin{figure}
\flushleft
\begin{tabular}{cc}
\includegraphics[width=13cm, height= 9 cm]{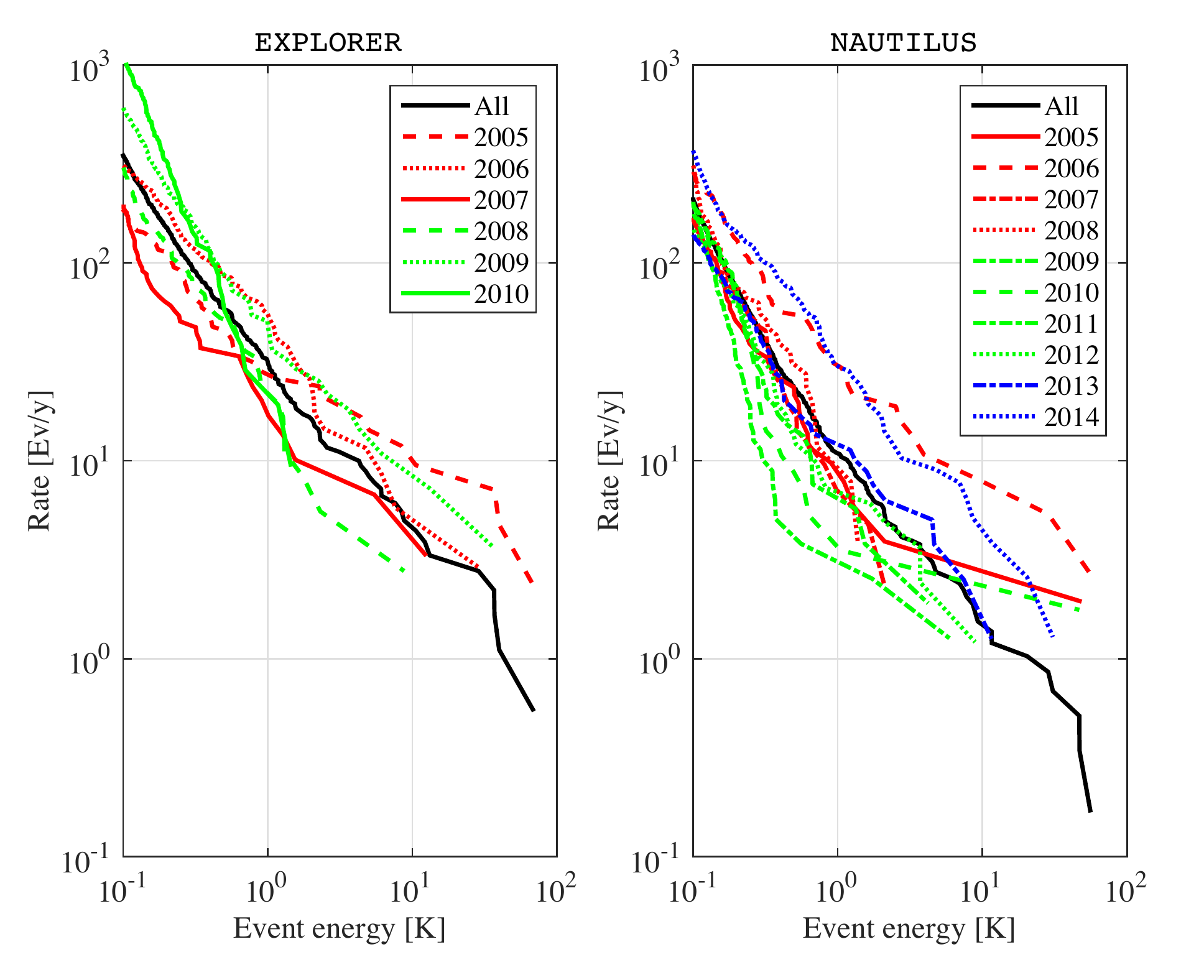}&

\end{tabular}
\caption{Rate of events of energy larger than the abscissa value
for Explorer (left) and Nautilus (right). The different lines are
the results at each different year, the overall result is shown
with linespoints.}
\label{fig:rate}
\end{figure}

In sect.\ref{proc2} we explained the procedures used to reject single
events considered incompatible with the kind of excitation we are
searching. Here we explain how we evaluated the ``false dismissal
probability" of 
these methods,  that is the probability
that we reject a good delta-like event. In order to evaluate this
probability, the procedure calls for the following steps:
\begin{itemize}
\item  randomly choose N time stamps inside the good operation periods
and extract from the filtered data a time segment around those
times
\item  inject in those time segments a copy of a real good event
(like for example the ones of fig.\ref{fig:ex_cosm} or
\ref{fig:na_cosm}), scaled to a given value of energy, repeating
this for different energies
\item  apply to this new time segments (data+event) the selection
procedures used to identify bad events and count how many we would
reject, at any value of injected energy
\item
finally,  determine the fraction of  injected events that were rejected  as a function of energy and assume that figure 
as the probability to reject a real, good event. 
\end{itemize}
The results are
summarized in tab.\ref{tab_fd} and we notice that the false
dismissal starts from $8 \%$ at the low energy edge for Explorer (100 injected signals in the 2008 data)
and from $5 \%$ for Nautilus (200 injected signal in the 2007 and 2014 data).
Then the false dismissal probability quickly drops, and is zero
at all energies above 0.25 K for both detectors.

\begin{table}
\centering
\begin{tabular}{||c||c|c||}
\hline\hline
 & \multicolumn{2}{c||}{False dismissal}\\
  Energy [K] & {\bf Explorer} &  {\bf Nautilus} \\
\hline\hline
 & & \\[-3mm]
 0.1   & 0.08  &  0.05  \\
 0.15  & 0.01  &  0.01  \\
 0.25  & 0.01  &  0.0   \\
 0.5   & 0.0   &  0.0   \\
\hline\hline
\end{tabular}
\vskip 0.1 in
\caption{ False dismissal values used in the analysis (see sect. \ref{proc4}). They drop to zero, in both detectors, for any signal energy larger than 0.25 K.}
\label{tab_fd}
\end{table}

We applied a similar procedure (injection of copies of a real
event at random times for a number of different energies) also
to determine the efficiency of detection. In this case we choose
1000 time segments, excluding the segments where an event over
threshold was present. We injected copies of a real event scaled
at 15 different energy values, ranging from 10 to 200 mK,
and counted how many of these events were {detected above} the 0.1 mK threshold.
The results are shown in fig.\ref{effic}.

\begin{figure}[h!]
\centering
\includegraphics[width=12.5cm,height=9cm]{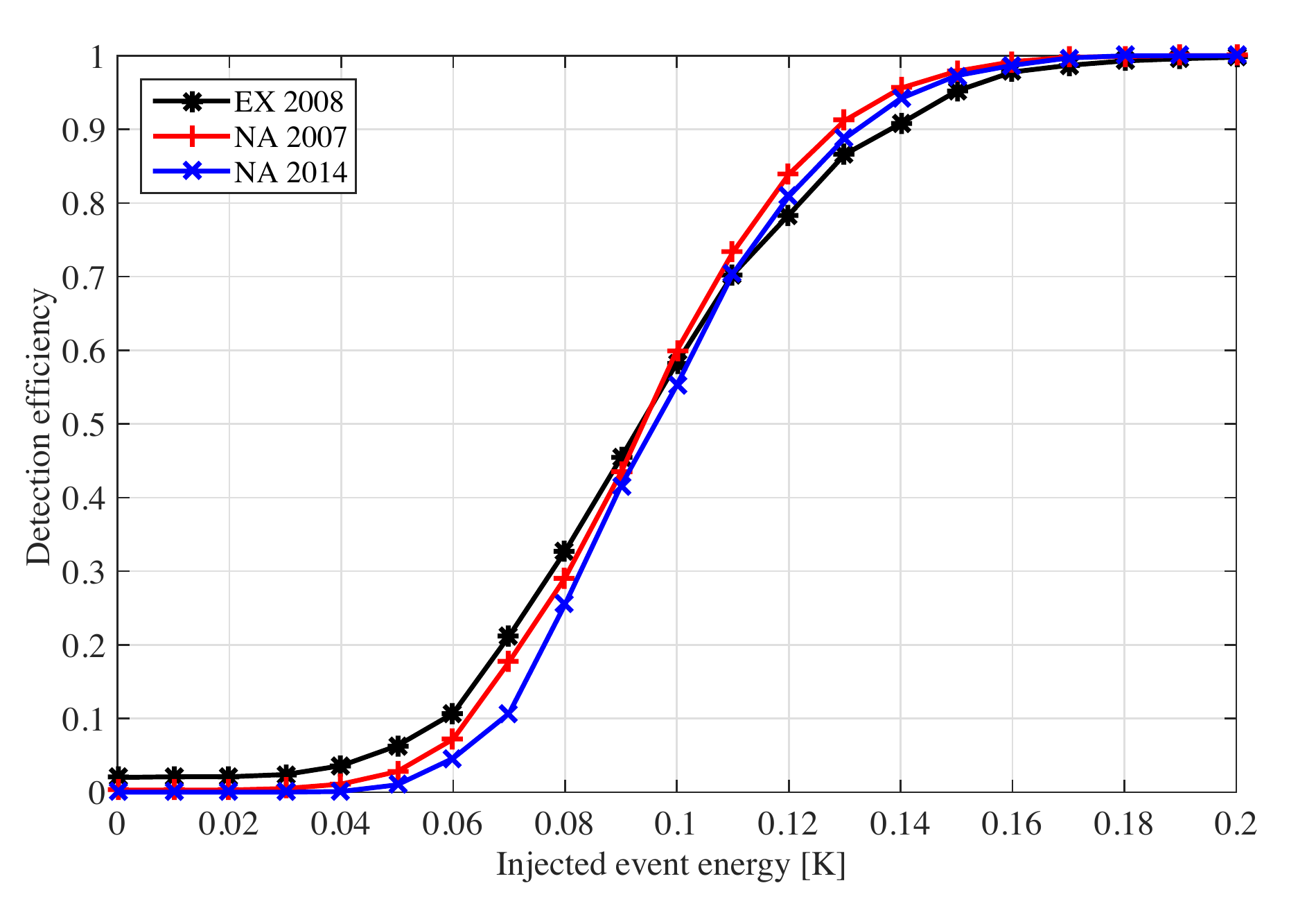}
\caption{Detection efficiency in three data sets}  
\label{effic}
\end{figure}

\section{Results}
\subsection{Nautilus}
The  cumulative  energy distribution  of the Nautilus events {surviving} the cuts described in the previous
section is shown in  fig.\ref{fig:rate} . The total livetime is 2129.1 days. This figure shows
that the shape of the distribution changes from one  year to another. For example the average value of the event energy is  0.2 $\pm$0.04 K in our ``best" year 2011, and 1.24 $\pm$ 0.58 K  in our ``worst" year 2006. 
This suggests that the operating conditions were not stationary and that there was  some unknown
source of  noise. A serious effort has been carried out to better understand  these possible noise sources
not considered in the previous paragraph;  in particular we have studied the correlation of the noise
events with the meteorological condition. We  found a weak correlation with the wind speed (or
the atmospheric pressure variations).  But the correlation is too feeble to proceed to a further data
selection. \\
{As an example of the calculations leading to upper limits, we show in
Fig.\ref{fig:Econfronto}  the results of  Montecarlo simulations for particles of a given mass and velocity: 
a newtorite of $M/v$=10 kg~s/km and  a nuclearite of
$\beta \theta(M)$=0.001.  The two energy distributions are compared with the experimental one.}

Note that the energy distribution of the nuclearite shown in this figure is quite
different from the data distribution. 

\begin{figure}
 \begin{center}
  \includegraphics[height=70mm, width=110mm]{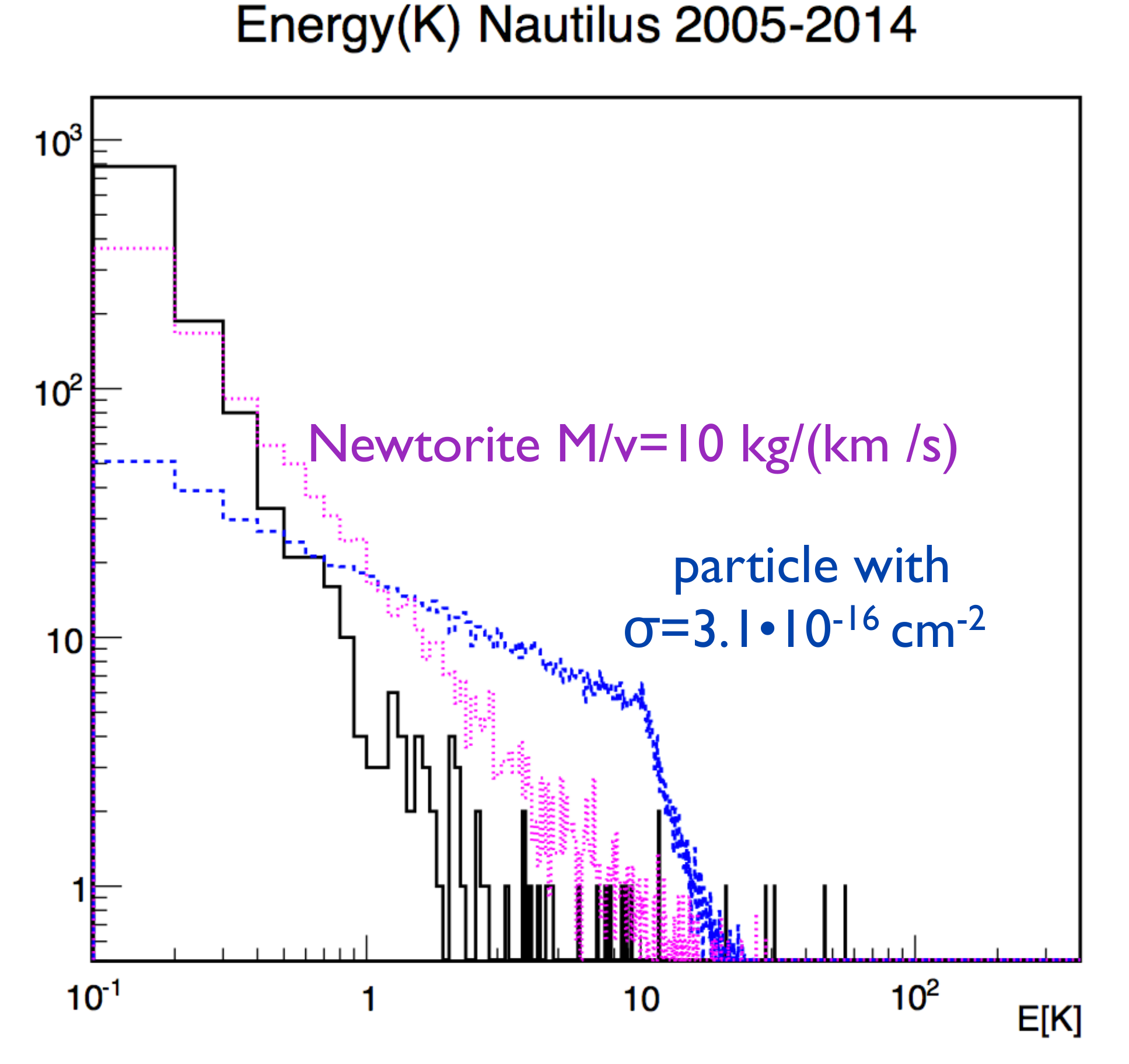}
  \end{center}
  \vspace{-0.5cm}
  \caption{Nautilus 2005-2014 events energy distribution (continuous line) compared with the nuclearite Montecarlo with $\beta\theta(M)$=0.001  (dashed, blue online) and with the newtorite
Montecarlo for $M/v$=10 kg~s/km (dots, magenta online). For different values of  $M/v$, eq.(\ref{eq:Tmiddle}) shows that the mode energy scales with $(M/v)^2$ Montecarlo events are normalized in order to have the same number of events as the real data.}
  \label{fig:Econfronto}
\end{figure}

\begin{table}[t]
\centering
\begin{tabular}{|c	|c	|c	|c	|c |c|}
\hline
 
$\beta \theta(M)$  & efficiency 	& $\sigma$  & events    &    flux upper limit  \\
 				&  	&($cm^2)$   & upper limit  &   ($cm^{-2} s^{-1} sr^{-1}$) \\
 
 \hline
0.0003               &  0.086    &$3.1 \cdot 10^{-16}$	& 825                          & 2.7
$\cdot 10^{-10}$ \\
0.0004               &  0.43   	&$3.1 \cdot 10^{-16}$	& 727                          &4.7
$\cdot 10^{-11 }$ \\
0.0005               &  0.61      &$3.1 \cdot 10^{-16}$	& 321                 		&1.4 $\cdot
10^{-11}$ \\
0.0008               & 0.81   	&$3.1 \cdot 10^{-16}$	&74                          	&2.6
$\cdot 10^{-12}$ \\
0.001                 &  0.83  	&$3.1 \cdot 10^{-16}$	& 52                       	& 1.7
$\cdot 10^{-12}$  \\
0.002                 &  0.92  	&$1.25 \cdot 10^{-15}$	&9.4                     	&  2.9
$\cdot 10^{-13}$\\
0.004                 &  0.96  	&${5.0} \cdot 10^{-15}$  	&3.2                		&  9.4
$\cdot 10^{-14}$\\
0.01                   & 0.97 	&$3.1\cdot 10^{-14}$  	&2.5                    	& 7.2 $\cdot
10^{-14}$\\
0.02                   & 0.99   	&$1.25\cdot 10^{-13}$  	& 2.4                      	& 6.8 
$\cdot 10^{-14}$\\
\hline
\end{tabular}
\caption{  Maximum number of nuclearites (or MACRO) events, compatible with the Nautilus energy distribution. Here, and in all following tables, the Upper Limits are calculated at the 90\%  Confidence Level (C.L.).  These values are computed using the optimal interval method.  The livetime is 2129.1 days.  In the case of nuclearite or of a generic MACRO the shape of the energy
distribution depends only on the geometry. The efficiency is defined as the ratio between the number
of events detected with the analysis cuts and the number of events hitting the bar. The
corresponding cross section $\sigma$ is for a typical DM velocity $\beta=0.001$. 
The flux is considered isotropic when computing upper limits.}
\label{tb1}
\end{table}

\begin{table}[t]
\centering
\begin{tabular}{|c	|c	|c	|c |c|}
\hline
 
$\beta \theta(M)$  & efficiency 	  & events   &    flux upper limit  \\
 				&  	   & upper limit  &   ($cm^{-2} s^{-1} sr^{-1}$) \\
 
 \hline
0.0003               &  0.086    		& 150                         & 3.6$\cdot 10^{-10}$ \\
0.0004               &  0.43   		& 55.5                          &2.7 $\cdot 10^{-11 }$ \\
0.0005               &  0.61      		&14.3                 		&4.7 $\cdot 10^{-12}$ \\
0.0008               & 0.81   		&8.54                       	&2.2 $\cdot 10^{-12}$ \\
0.001                 &  0.83		& 7.97                        	& {2.0} $\cdot 10^{-12}$  \\
0.002                 &  0.92  		&{3.30}                         & 7.4 $\cdot 10^{-13}$\\
0.004                 &  0.96  	 	&2.62                 		&  5.6 $\cdot 10^{-13}$\\
0.01                   & 0.97 	  	&{2.40}                    	& 5.1 $\cdot 10^{-13}$\\
0.02                   & 0.99   	  	& 2.35                      	& 4.9  $\cdot 10^{-13}$\\
\hline
\end{tabular}
\caption{ Nuclearites upper limits as in Tab.\ref{tb1} but for Nautilus 2011, the ``best" data
set. The  livetime is 287.8 days. This data set is the one with the lowest noise. The flux upper limits
are better than with the full set of data when dominated by the noise, but are worst at  $\beta \theta(M)  \gtrsim $0.001 when they are
dominated by the statistics}
\label{tb2}
\end{table}

For the signal energy distribution we assume that  computed by the Montecarlo. Upper limits are
computed for different values of $\beta \theta(M)$ and $M/v$.  
{For both the nuclearite (or Macro) and the newtorite cases, }
the energy  distribution only depends  on the geometry, and it is well predicted.

To compute the limits  on the maximum  allowed  number of events for both nuclearites and newtorites, we have used  the so called {\it optimum interval method} to find
an upper limit for a one-dimensionally distributed signal in the presence of an unknown background
\cite{Yellin:2002xd, Yellin:2008da, software}.

Tab.\ref{tb1} shows  the Nautilus 2005-2014   90\%  C.L. upper limits on
the number of possible nuclearites in the data set, and the resulting
limit on the flux. The limits are for an
isotropic  flux, not considering possible absorption effects by the Earth or the
atmosphere (see below). The efficiency is defined as the ratio between the number of events detected
with the analysis cuts and the number of events hitting the bar.

Tab.\ref{tb2} show  the same kind of limits but for  Nautilus 2011 only, the year with the lowest
noise. The flux upper limits are better than those  of the complete data set when dominated by the
noise, and are worse at  $\beta \theta(M)  \gtrsim $0.001 when they are dominated by the statistics.

Finally,  the same upper limits calculation is applied to the case of newtorites,
comparing the energy distribution of data and the newtorites Montecarlo. 
In this case the energy
distributions do not  change much with the parameter $M/v$: this is because, as $M/v$ increases,
the sensitive volume also increases, allowing detection of signals from newtorites passing at larger distances from the bar. Therefore
the upper limits for the number of events is almost insensitive on $M/v$. 
In this Montecarlo,  newtorites are extracted on a cylindrical  surface much larger than the
antenna bar. The dimension of the extraction surface is optimized for different $M/v$ ratios (larger for
bigger $M/v$). The Montecarlo therefore computes the efficiency in the case of simulated events that release at least 0.1 K and survive the analysis cuts. 
For example, the acceptance  for the simulation of newtorites of $M/v$=10 kg~s/km is  426 $m^2sr$
(including the efficiencies due to the data selection), while
for $M/v$=60 kg~s/km the acceptance    {grows to}  2398 $m^2sr$. The signal  energy increases as $(M/v)^2$,
this produces a linear increase of the acceptance with $M/v$.

The results are shown in Tab.\ref{tb3}. For sake of uniformity, we have  computed the newtorites upper limits 
using the same optimal interval method as in the nuclearite case. But in this case this tool
does not help very much since the shape of the expected energy distribution {
closely resembles that of the data.  As a consequence, it is }more important to select data with very low
noise than to increase the livetime. In Tab.~\ref{tb6}  we can see that the limits obtained  using the Nautilus 2011
data are better than those obtained using the full Nautilus data set in almost  the entire $M/v$ range.
\begin{table}[t]
\centering
\begin{tabular}{|c	|c |c	|c	|c }
\hline
 
M/v  &  acceptance 	&   events   &  flux upper  limit\\
  ($kg~ km^{-1}~ s$)      & ($m^2 sr$) 	&  upper limit   &  ($cm^{-2} s^{-1} sr^{-1}$) \\
 \hline
1               & 33.4 	&545   	        &8.9 $\cdot 10^{-12 }$ \\
2               & 85.3 	&365   	        &2.4 $\cdot 10^{-12}$ \\
5               & 209	&253                &6.5 $\cdot 10^{-13}$ \\
10             &  426	&257	       & 3.5 $\cdot 10^{-13}$  \\
20             &  888		&244	       & 1.5 $\cdot 10^{-13}$\\
40             &  1652	&248	       & 8.3 $\cdot 10^{-14}$\\
60             &  2398	&242 	      & 5.5 $\cdot 10^{-14}$\\
\hline
\end{tabular}
\caption{Nautilus 2005-2014, newtorites upper limits. Note the linear increase of the acceptance
with $M/v$. }
\label{tb3}
\end{table}

\begin{table}
\centering
\begin{tabular}{|c	|c	|c	|c }
\hline
 
M/v  & 	    events   &   flux upper  limit \\
  $kg~ km^{-1}~ s$      &  	  upper limit   &  $(cm^{-2} s^{-1} sr^{-1)}$ \\
 \hline
1               &  37   	        &4.5 $\cdot 10^{-12 }$ \\
2               &  23   	        &1.1 $\cdot 10^{-12}$ \\
5               & 22	                &4.3 $\cdot 10^{-13}$ \\
10             & 22	                & 2.1 $\cdot 10^{-13}$  \\
20             &22	       		& 1.0 $\cdot 10^{-13}$\\
40             & 22	       		& 5.5 $\cdot 10^{-14}$\\
60             & 21	     	 	& 3.5 $\cdot 10^{-14}$\\
\hline
\end{tabular}
\caption{Nautilus 2011. newtorites upper limits. Livetime=278.8 days, 160 events $\ge$  0.1K}
\label{tb6}
\end{table}

\subsection{Explorer}
The Explorer data are selected according to a  procedure similar to  that of Nautilus. The  energy distribution  of the events in Explorer,
starting from 2005, is shown in fig.\ref{fig:rate}. 
The total livetime is 658.56  days, much shorter than for Nautilus, also because it ceased operation
in June 2010.
This figure shows that the shape of the distribution changes from year to year, similarly to what occurs
in Nautilus. The corresponding nuclearite upper limits are reported in Tab~\ref{tb4}. Explorer and Nautilus
are very similar detectors, but there are some differences: e.g. in the geographical location, or
in the cosmic ray detectors used as veto. So, it could be interesting to separately analyze their limits 
before combining the data. 

\begin{table}[t]
\centering
\begin{tabular}{|c	|c	|c	|c |c|}
\hline
 
$\beta \theta(M)$  & efficiency 	  & events    &     flux upper limit  \\
 				&  	   & upper limit  &   $(cm^{-2} s^{-1} sr^{-1)}$ \\
 
 \hline
0.0003               &  0.086    	& 498                         & 5.2 $\cdot 10^{-10}$ \\
0.0004               &  0.43   		& 409                         &8.6 $\cdot 10^{-11 }$ \\
0.0005               &  0.61      	& 221               		     &3.2 $\cdot 10^{-11}$ \\
0.0008               & 0.81   		& 61                       	&6.9 $\cdot 10^{-12}$ \\
0.001                 &  0.83  		& 48                        	& 5.2 $\cdot 10^{-12}$  \\
0.002                 &  0.92  		& 11.2                       & 1.1 $\cdot 10^{-12}$\\
0.004                 &  0.96  	 	& {3.30}                 		&  3.1 $\cdot 10^{-13}$\\
0.01                   & 0.97 	  	& 2.55                    	& 2.4 $\cdot 10^{-13}$\\
0.02                   & 0.99   	  	& 2.43                      	& 2.2  $\cdot 10^{-13}$\\
\hline
\end{tabular}
\caption{ Nuclearites upper limits as in Tab.\ref{tb1} but for Explorer from 2005 until 2010. 
The  livetime is 658.5 days.  The flux upper limits  at  $\beta \theta(M)  \gtrsim $ 0.004   are
dominated by statistics. }
\label{tb4}
\end{table}

\subsection{Nautilus+Explorer}
Adding the Explorer  noisier data to the Nautilus data degrades the upper limit for low values of
$\beta \theta(M)$, where the limit is due to the noise.
This data combination is only useful  at large values of $\beta \theta(M)$. In Tab.~\ref{tb8} we show the
limits for this data combination for nuclearites. 

\begin{table}[t]
\centering
\begin{tabular}{|c	|c	|c	|c	|c |c|}
\hline
 
$\beta \theta(M)$  & efficiency 	& $\sigma$  &  events  &     flux upper limit  \\
 				&  	& ($cm^2$)  & upper limit  &  ($cm^{-2} s^{-1} sr^{-1}$) \\
 
 \hline
0.0003               &  0.086    &$3.1 \cdot 10^{-16}$	& 848                          & {2.1}
$\cdot 10^{-10}$ \\
0.0004               &  0.43   	&$3.1 \cdot 10^{-16}$	& 1077                          & {5.3}
$\cdot 10^{-11 }$ \\
0.0005               &  0.61      &$3.1 \cdot 10^{-16}$	& 511                 		&1.76
$\cdot 10^{-11}$ \\
0.0008               & 0.81   	&$3.1 \cdot 10^{-16}$	&112                          	&2.95
$\cdot 10^{-12}$ \\
0.001                 &  0.83  	&$3.1 \cdot 10^{-16}$	& 84                       	& 2.13
$\cdot 10^{-12}$  \\
0.002                 &  0.92  	&$1.25 \cdot 10^{-15}$	& 11.2                     	&  2.58
$\cdot 10^{-13}$\\
0.004                 &  0.96  	&$5 \cdot 10^{-15}$  	&3.32                		&  7.37
$\cdot 10^{-14}$\\
0.01                   & 0.97 	&$3.1\cdot 10^{-14}$  	&2.56                    	& 5.58
$\cdot 10^{-14}$\\
0.02                   & 0.99   	&$1.25\cdot 10^{-13}$  	& 2.44                      	&
5.23 $\cdot 10^{-14}$\\
\hline
\end{tabular}
\caption{ Nuclearites (or MACROs) 90\%  C.L. maximum number of events compatible with the  energy
distribution of the full data set Nautilus+Explorer, computed using the optimal interval method. The
livetime is 2787.6 days.  All other considerations detailed in table \ref{tb1} apply here.}
\label{tb8}
\end{table}

\begin{figure}[h]
\begin{center}
\includegraphics[width=10cm]{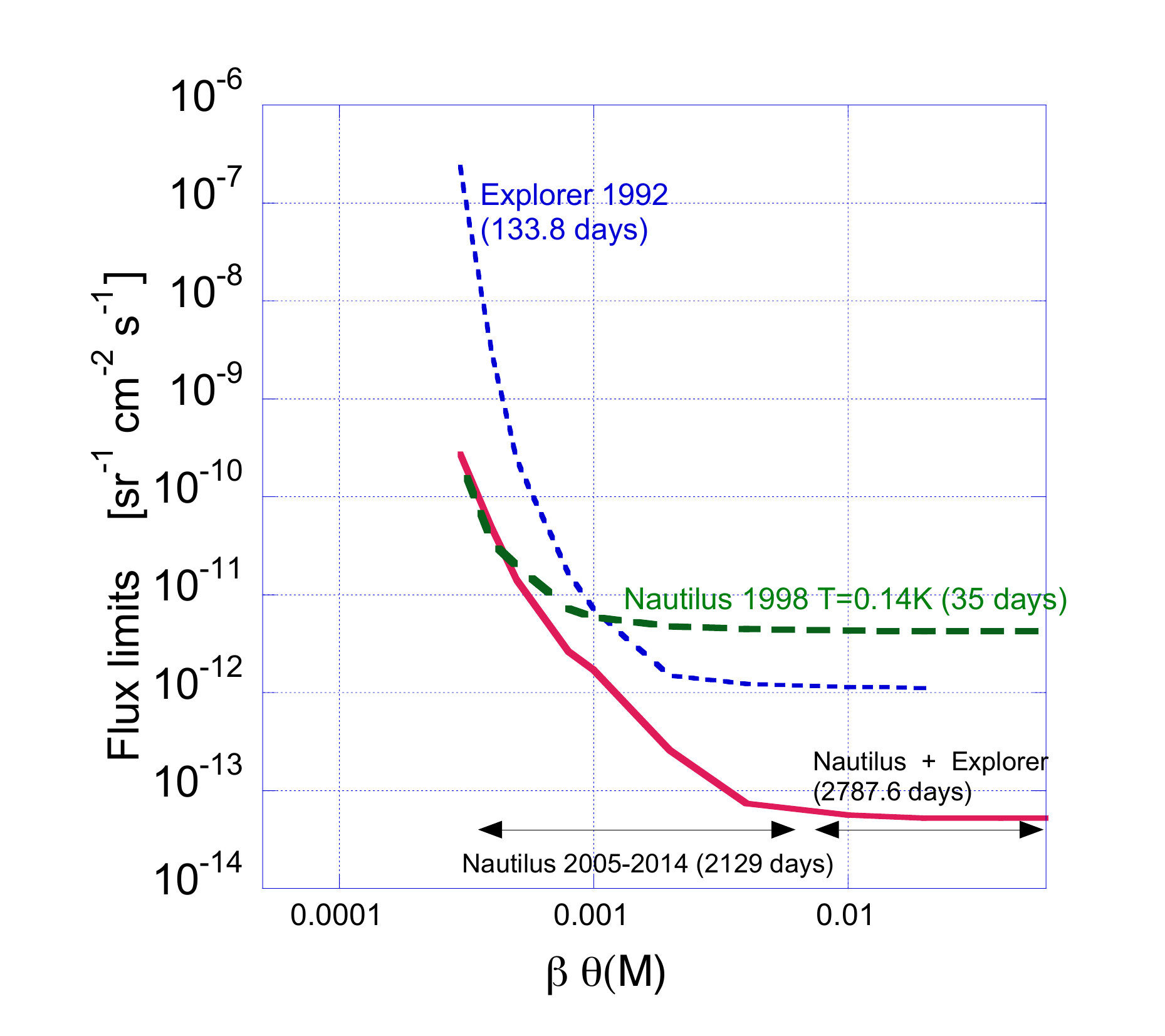}
\caption{ 90\% C.L. upper limits  for an  isotropic  flux of nuclearites, compared to  previous results (Explorer)
\cite{explorer}.
The continuous line shows the results of this analysis. The Explorer data are used in combination to the
Nautilus data only for $\beta\theta(M)>0.002 $.
For nuclearites that cannot penetrate the Earth, the flux limit should be doubled.
 Limits  from a short run with Nautilus at $T=0.14$K, long dashed line, may be interesting because
of the different detection mechanism  in the superconducting state.}
\label{limiti}
\end{center}
\end{figure}

\section{Discussion and dark matter limits}
\subsection{Nuclearites}
The nuclearite flux upper limits are summarized in fig.\ref{limiti}.
This figure  also shows the limits  from a short run with Nautilus at $T=0.14$K, with live-time=
35.1 days.  This result is interesting because of the different detection mechanism  in the
superconducting state.
For   $\beta\theta(M)>0.01$, where the background is negligible, the flux upper limit is dominated
only by the live-time. Note that in this search the events in coincidence with the cosmic ray {shower}
detector are removed from the Nautilus data.  {In principle,  fast nuclearites could produce light in the Explorer
scintillators (due to black body emission\cite{derujula}), light  that could be confused with a cosmic ray
event. Therefore we have verified the ``shower" signature of  all cosmic rays in  the Explorer data: they indeed hit all
scintillator counters as expected from a big  extensive air  shower.}

{In order to compare our results with previous searches, we recall that} many techniques
 have been used to detect
nuclearites:  damages in plastic
materials like CR39, Makrofol or Lexan, light emission in oil or sea
water \cite{MACRO}, \cite{ANTARES}, seismic waves induced by big nuclearites. Due to the
uncertainties in the conversion of the energy losses in a measurable signal it is important that
different
techniques are used to detect such exotic particles. 

The best limits   above sea level are obtained with track etch detectors, but the detection
mechanism is more complicated than the ``calibrated" calorimetric technique used in this search.
 The SLIM limit\cite{SLIM} for $ \beta=10^{-3}$ is   $1.3 \cdot 10^{-15}  cm^{-2} s^{-1} sr^{-1}$.
and the OHYA\cite{Orito:1990ny} limit is $3.2 \cdot 10^{-16}  cm^{-2} s^{-1} sr^{-1}$, both stated with a $90\%$ confidence limit. The limits
from track etch in old mica\cite{Price:1988ge}  depend on several additional
assumptions. 
In general we observe that there is no quantitative theory describing  the track etch mechanism.
Track etch detectors have been calibrated with slow charged ions, assuming energy lost  by Coulomb
elastic collisions.  In principle this process is different from the energy loss of eq.\ref{eq:rel}.
The extrapolation of those calibrations to neutral massive particles, losing energy by atomic
collisions, is not straightforward.
The {unique feature} of our search is  the calorimetric technique  that directly measures the energy loss, 
and therefore is less sensitive to uncertainties or hypotheses in the detection mechanism.

 Finally figure~\ref{limitimassa}  shows the upper limits vs  the nuclearite mass. For  mass between
 $5\cdot 10^{-14}~g$ (threshold due to the atmosphere) and $10^{-4}g$ this limit is significantly
smaller than the flux of galactic dark matter for $\beta=10^{-3}$ and $\rho_{DM} =0.3~GeV/cm^3$. 
Although the upper limit of fig.(\ref{limitimassa}) is computed for a set  velocity,  a sample simulation for $M \le 1.5~ ng$  shows that the use of a Maxwell-Boltzmann distribution of velocities, centered at  $\beta=10^{-3}$,  produces an upper limit consistent (6\% larger)  with that reported here.

\begin{figure}[h]
\begin{center}
\includegraphics[width=12cm, height=9cm]{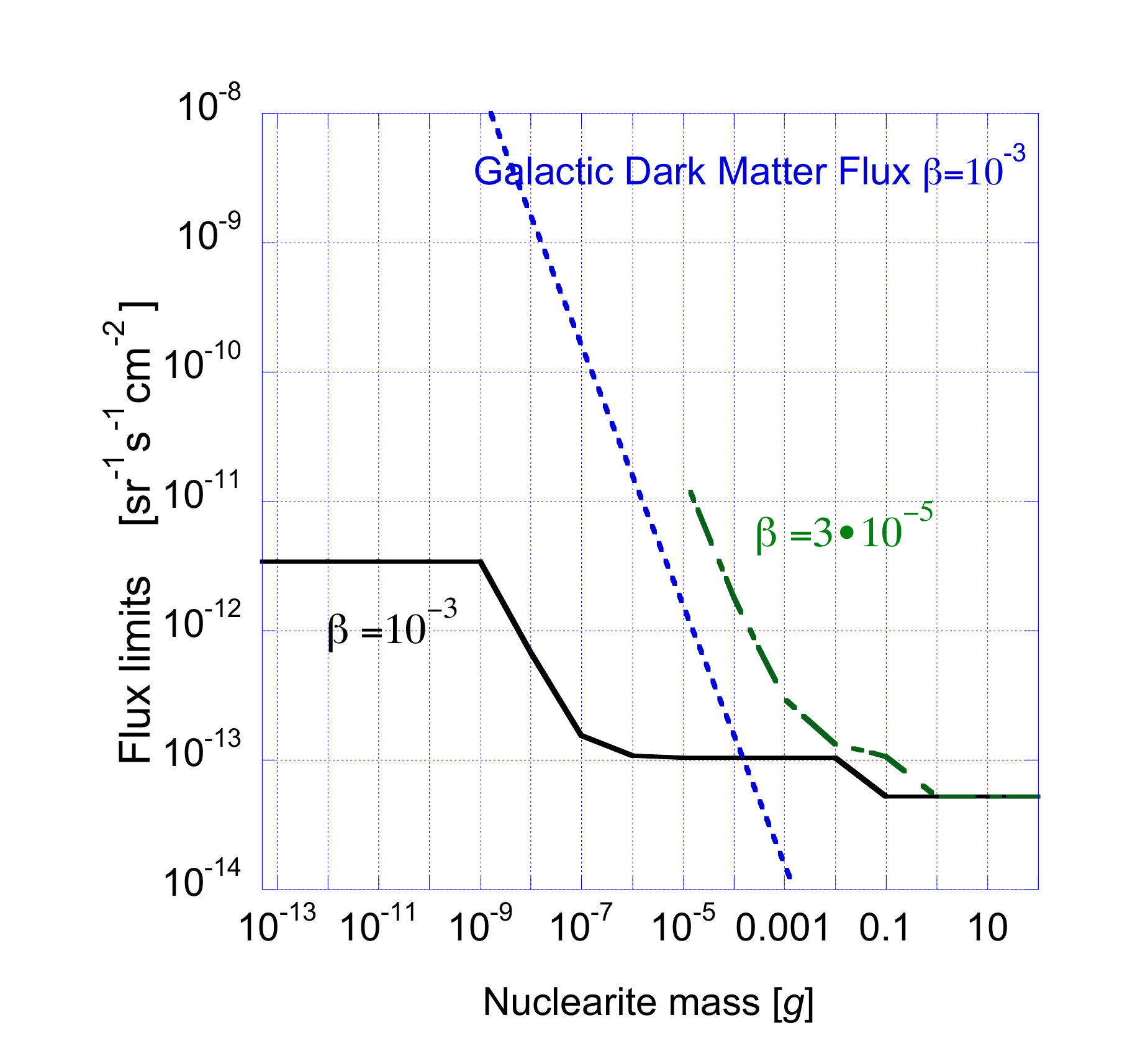}
\caption{ Flux upper limits for nuclearites with $\beta=10^{-3}$ and  $\beta=3\cdot 10^{-5}$  (Earth escape velocity)
vs  mass. The limits are derived from  Fig~\ref{limiti}, computing the appropriate
$\beta\theta(M)$. For some combination of masses and $\beta$, nuclearites cannot cross the earth: in
this case a factor 2 is applied to the  limits of fig.\ref{limiti}. For  mass between  $5\cdot
10^{-14}~g$ (threshold due to the atmosphere) and $10^{-5}g$ this limit is significantly smaller
than the flux of galactic dark matter for $\beta=10^{-3}$. }
\label{limitimassa}
\end{center}
\end{figure}

\subsection{MACROs}
Nuclearites are a particular kind of MACRO with a specific characteristic: it 
 is supposed to have
dimension and cross section  always larger than $3.1\cdot10^{-16} cm^2$ while a generic MACRO can
be smaller.  This could have some important experimental implications for track etch
mica or plastic detectors.  As observed in \cite{Jacobs:2014yca}, the requirement for an etchable
track is for the burrowed hole in the mica sample to be large enough that hydrofluoric
acid can penetrate it during the etching process. This is plausible for hole diameters larger than
a few tens of nanometers. Even considering that $\delta$ rays (secondary electrons of a few $keV$) could contribute to enlarge the hole, 
it is worth noting that the calibrations of  track etch detectors are done with charged ions and
therefore with object of the size of 
a few tens of nanometers.
This consideration does not apply to $gw$ bar detectors that  have no such limitation in the object
dimensions. 

 In crossing the atmosphere and the Earth, the deceleration given by eq.(\ref{eq:vel})  sets bounds on the flux of detectable MACROs as a function of  $\frac{\sigma}{M}$. For a down-going vertical track  the upper limit
is $\frac{\sigma}{M} =5.6\cdot10^{-3}$  $cm^2~g^{-1}$. For a vertical up-going MACRO crossing the
Earth the limit is $\frac{\sigma}{M} =7.8\cdot10^{-10}$$cm^2~g^{-1}$.
The upper limits on the cross sections shown in Fig~\ref{limiticross} have been computed using the
same procedure as for nuclearite. 

An allowed region for the MACRO dark matter can be obtained in the plane cross section  vs. mass, by
excluding the region with upper limit on the flux smaller than the dark matter flux following the
approach of Ref~\cite{Jacobs:2015csa}.
Fig.\ref{esclMACRO} shows the excluded region. It is interesting to observe that studies of
cosmological galaxy and cluster halos suggest a value $\sigma/M=0.1$ $cm^2$/g  \cite{Rocha:2012jg, Peter:2012jh}. For the comparison with constraints using other techniques see 
Ref.~\cite{Jacobs:2015csa}. Here, once again, we want to stress the difference of this detector with
respect to all other techniques. For small cross sections $\sigma \lesssim 5 \cdot 10^{-17} ~cm^2$ we have a
threshold due to the 0.1 K cut used in this analysis. This implies that the excluded area cannot extend
below this value. Finally, we remark that for very high cross sections,
$\sigma \gtrsim 10^{-14} ~cm^{-2}$, corresponding to signals  of the order of $10^4  K$,  the efficiency of our search is strongly reduced due to saturation of the data acquisition {electronics}.
\begin{figure}[h]
\begin{center}
\includegraphics[width=9cm]{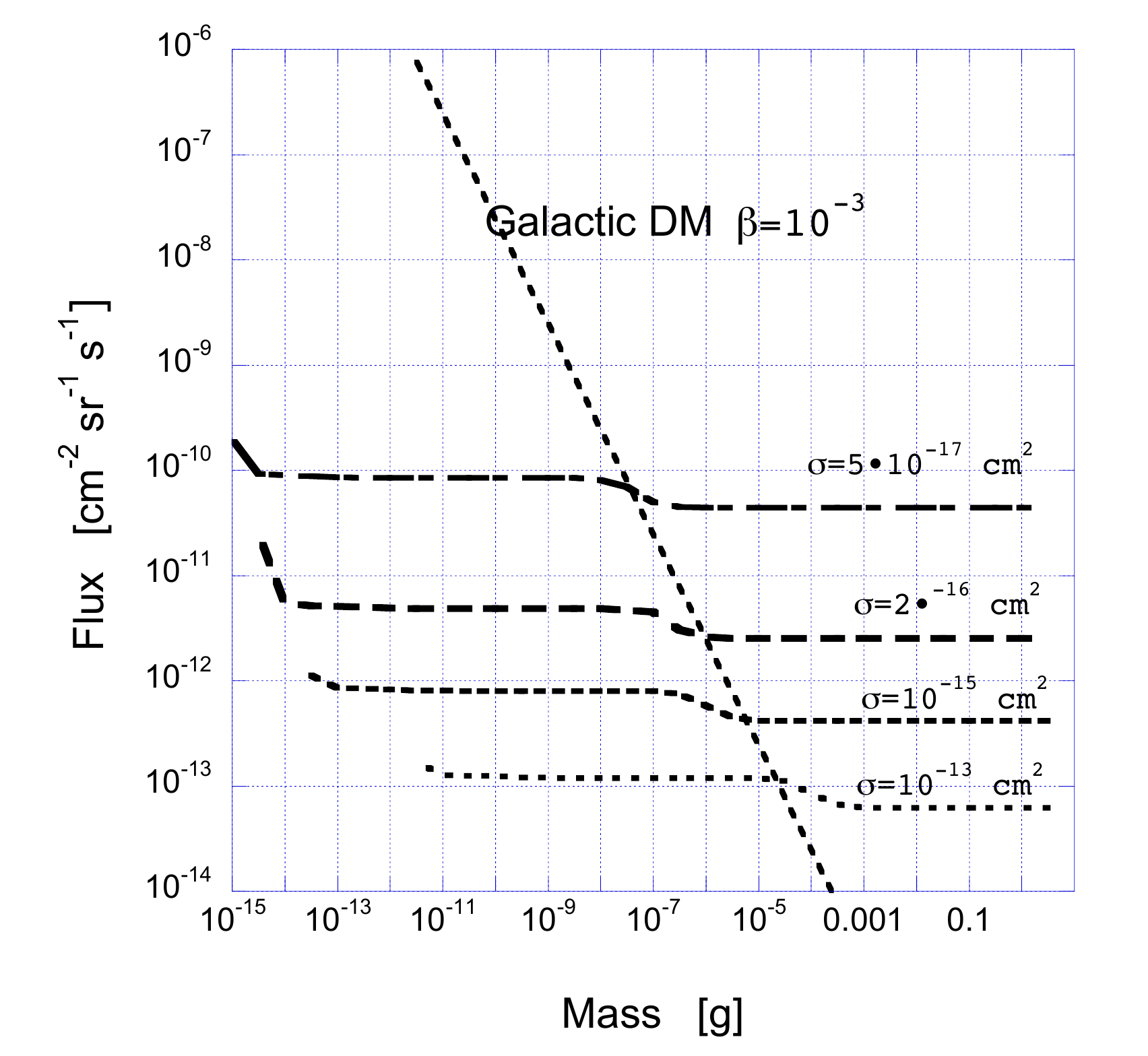}
\caption{ 90 \% CL flux upper limits  {vs MACRO mass} for different cross sections.  The lowest mass limit on abscissa is set by a mass large
enough to cross the atmosphere.}
\label{limiticross}
\end{center}
\end{figure}

\begin{figure}[h]
\begin{center}
\includegraphics[width=12cm]{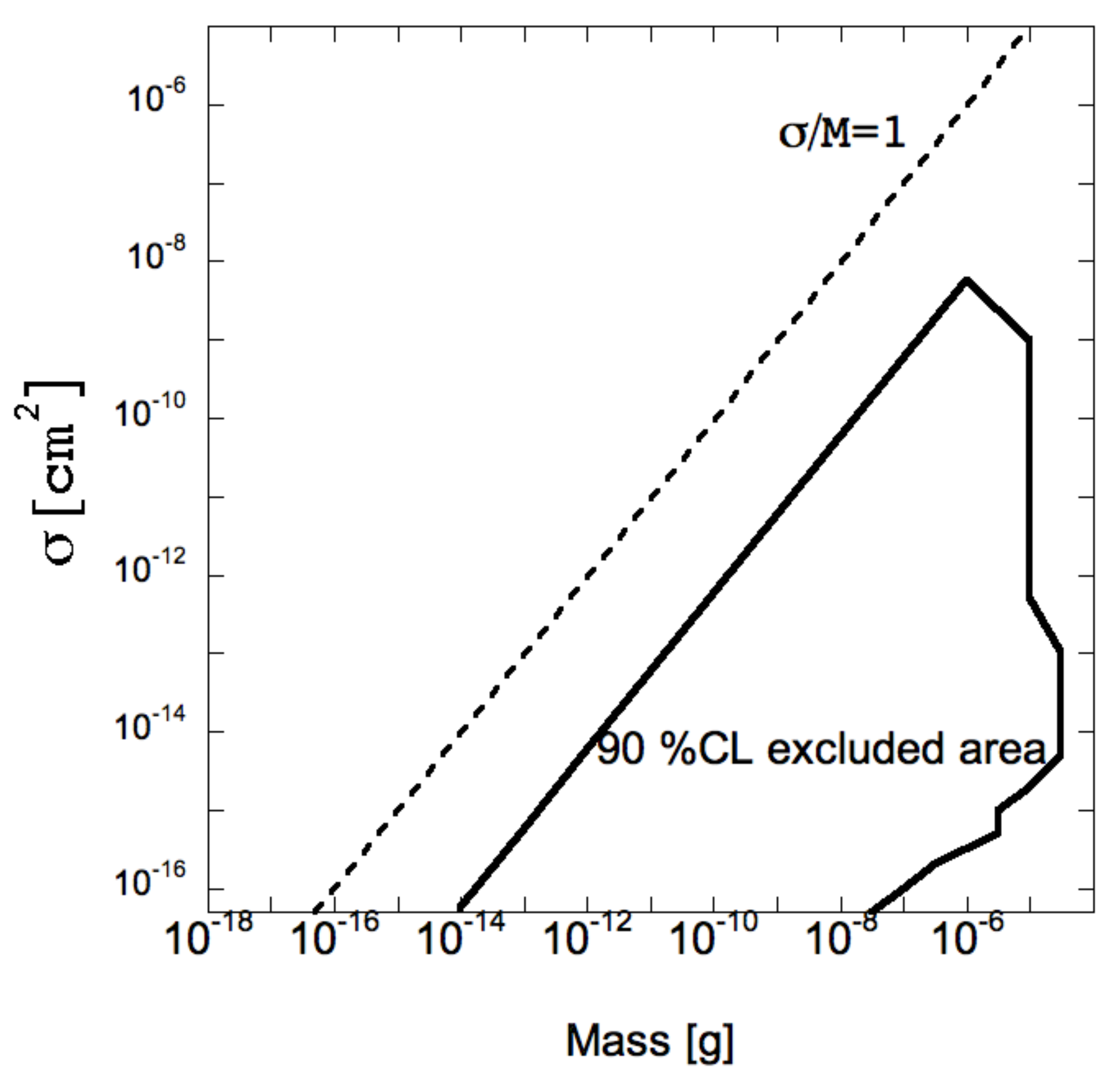}
\caption{ MACRO's 90 \% CL excluded regions in the plane cross section - mass. The  line $\sigma$/M=1 is drawn for reference.
The study of cosmological halos may suggest a value $\sigma$ /M=0.1.}
\label{esclMACRO}
\end{center}
\end{figure}

\subsection{Newtorites}
The limit of Tab.~\ref{tb6} are shown in fig.\ref{limitiNewt} for two values of the velocity $v$ =10
and  300 km/s. We have only used  the Nautilus 2011 data set  because, in this case, it is more
important to have low noise data. The limits reach a plateau as the mass increases. This
plateau could in principle extend to very large masses; however,  the use of the filter matched to delta - like events
limits the search to signals in the ms range and this sets a limitation on d/v where d is the average distance
from the bar and $v$is the particle velocity. 
In turn, this implies a limitation in the acceptance. 
Our limit, although very far from the DM expected density ($5\cdot10^{-13}~kg/km^3$),  could be of some
interest due to the lack of other experimental limits derived from the direct detection of DM
particles that only interact gravitationally and on the Earth. 

There are several limits obtained studying the motion of celestial body in the solar system.
For example Adler \cite{Adler:2008rq} obtains a  direct upper limit of the mass of Earth-bound dark
matter lying between the radius of the moon orbit and the geodetic satellite orbit. The value
obtained is $0.13~ kg/km^3$, larger than our limit shown in fig.\ref{limitiNewt}. 
Considering larger volumes  Pitjev \cite{Pitjev:2013sfa} has found a limit for possible DM inside
the Earth-Sun orbits of the order of $1.4\cdot10^{-7} kg/km^3$.

Our direct limit on newtorites  could be improved by orders of magnitude using two or more nearby
bar antennas in coincidence. The performances for newtorites of two antenna in coincidence has been
studied by a Montecarlo simulation that uses as input the Nautilus 2011 data set (therefore assuming the same
performances of Nautilus 2011).  
In the Montecarlo we  assumed  two antennas, positioned 1.5 m apart, with uncorrelated noise.  Larger distances, up to tens of meters, can still produce a detectable signal, depending on the value of $M/v$.
The result of this study is that a gain of about 300 seems to be possible
in 10 years of operations with noise similar to that of Nautilus 2011. This gain  is not enough
to reach the Pitjev bound. To reach this bound  it

is necessary to  increase the number of antennas and to reduce their noise. We recall that the current
antenna noise is limited by technology and is far from the intrinsic quantum limit of this kind of
device $\Delta E= \hbar\omega_{0}= 6~10^{-31}$ joules. But  a large R\&D effort would be necessary to
approach this limit.
\begin{figure}[h]
\begin{center}
\includegraphics[width=12cm]{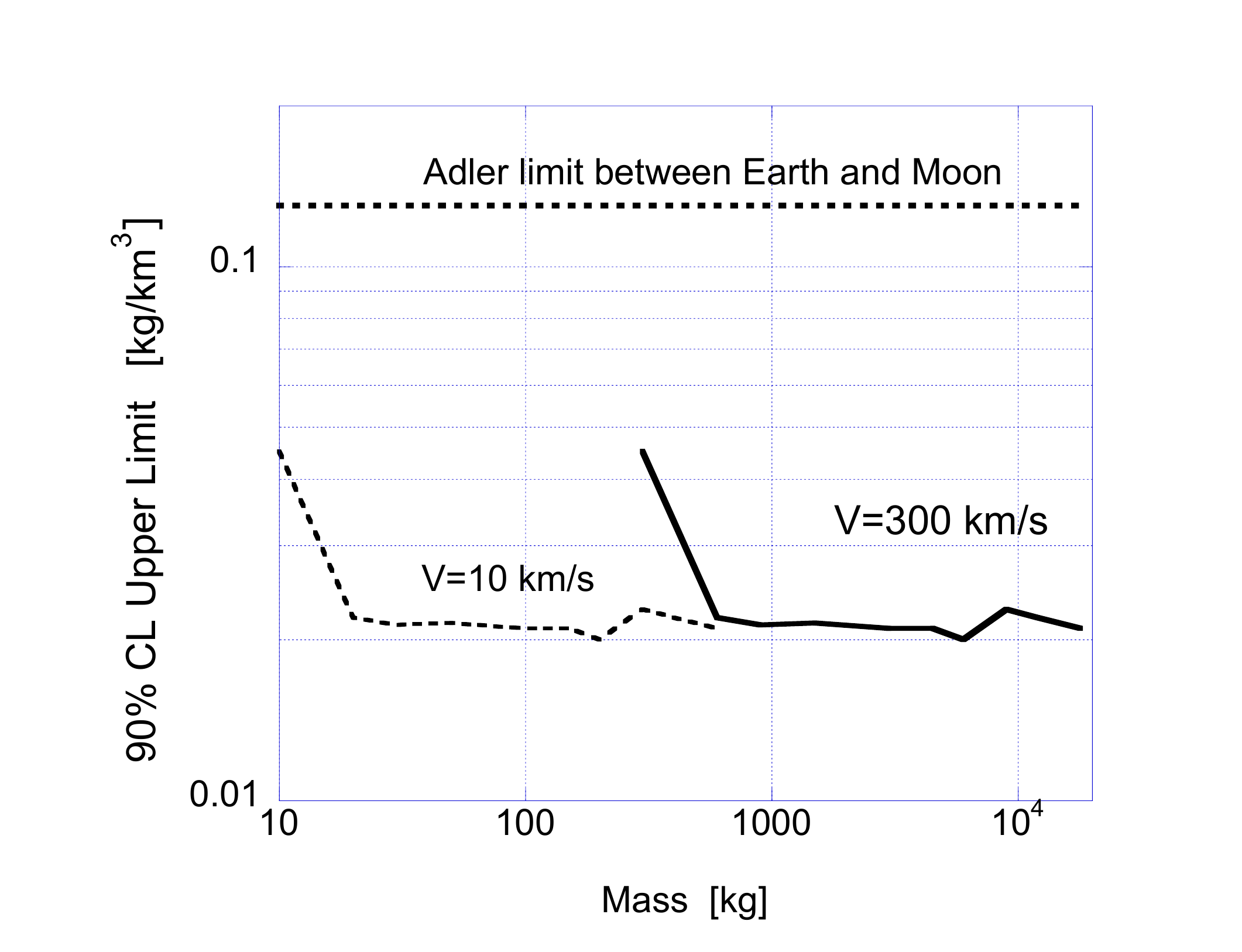}
\caption{ Newtorites density upper limits for v=10 and v=300 km/s vs  the newtorite
mass. The limits are obtained from the Nautilus 2011 data set. As the  mass increases the limit reaches a plateau  }
\label{limitiNewt}
\end{center}
\end{figure}

\begin{center} 

*** 
\end{center}

We gladly acknowledge precious help from our technicians M. Iannarelli, E. Turri, F. Campolungo, R.
Lenci, R. Simonetti and F. Tabacchioni. We are grateful to our unknown referee, for the careful reading and for many suggestions that prompted us to produce a better paper.

\bibliographystyle{model2-names}
\bibliography{<your-bib-database>}

\end{document}